\documentclass[%
 reprint,
 superscriptaddress,
 amsmath,amssymb,
 aps,
 pra,
floatfix
]{revtex4-2}
\pdfoutput=1
\usepackage[utf8]{inputenc}
\usepackage[english]{babel}
\usepackage[T1]{fontenc}
\usepackage{amsmath,amssymb}
\usepackage{hyperref}
\usepackage{braket}
\usepackage[qm]{qcircuit}
\usepackage{graphicx}
\usepackage[sort&compress]{natbib}

\usepackage{subcaption}

\usepackage{tikz}
\usepackage{lipsum}
\usepackage{csvsimple}
\usepackage{rotating}

\newcommand{\R}{\mathbb{R}}
\newcommand*\diff{\mathop{}\!\mathrm{d}}
\newcommand{\bs}{\boldsymbol}
\newcommand{\mc}{\mathcal}

\begin{document}

\title{A Performance Study of Variational Quantum Algorithms for Solving the Poisson Equation on a Quantum Computer}

\author{Mazen Ali}
\email{mazen.ali@itwm.fraunhofer.de}
\author{Matthias Kabel}
\email{matthias.kabel@itwm.fraunhofer.de}
\affiliation{Fraunhofer ITWM, 67663 Kaiserslautern, Germany}



\begin{abstract}
	Recent advances in quantum computing and their increased availability has led to a growing
	interest in possible applications.
	Among those is the solution of partial differential equations (PDEs) for, e.g.,
	material or flow simulation.
	Currently, the most promising route to useful deployment of quantum processors
	in the short to near term are so-called hybrid variational quantum algorithms (VQAs).
	Thus, variational methods for PDEs have been proposed as a candidate for quantum
	advantage in the noisy intermediate scale quantum (NISQ) era.
	In this work, we conduct an extensive study of utilizing VQAs on
	real quantum devices to solve the simplest prototype of a PDE
	-- the Poisson equation.
	Although results on noiseless simulators for small problem sizes may seem deceivingly
	promising, the performance on quantum computers is very poor.
	We argue that direct resolution of PDEs via an amplitude encoding of the solution is not
	a good use case within reach of today's quantum devices
	-- especially when considering large system sizes and
	more complicated non-linear PDEs
	that are required in order to be competitive with classical high-end solvers. 
\end{abstract}

\maketitle


\section{Introduction}
The technological progress in quantum computing has spurred
a lot of research into applications with the potential for an advantage
over classical computing.
One of these possible applications is the solution of partial differential
equations (PDEs) that are extensively used in various areas of engineering such
as computational fluid dynamics (CFD) \cite{Iliev2004}
or material simulation \cite{ANDRA201333, Kabel2014}.

In its simplest form, a linear PDE is transformed via a discretization method into
a system of linear equations. The latter can then be solved with the quantum HHL
method \cite{hhl}. For general linear PDEs in three spatial dimensions,
one can expect at best a quadratic speedup compared to classical
solvers, see \cite{hhl_fem}.
The speedup may, however, increase for high-dimensional PDEs.
Despite recent progress with HHL \cite{Saha22, Robson22, Vazquez22}, it requires
deep entangling circuits and has thus limited scalability within
the noisy intermediate scale quantum (NISQ) era computers.

A more NISQ friendly alternative are so-called \emph{(hybrid) variational algorithms}.
These are (mostly heuristic) methods that rely on a quantum-classical
approach where the quantum computer is only used to execute relatively
shallow circuits to estimate cost functionals within a classical optimization loop.
Hybrid methods have gained a lot of attention \cite{Cerezo21} as \emph{the}
class of methods for NISQ devices, including applications to PDEs
\cite{Sato21, Liu21, Demirdjian22}.

The simplest example of a model PDE problem is the 2nd order linear Poisson equation.
This has been previously addressed in \cite{Sato21, Liu21} using the hybrid variational quantum linear
solver (VQLS) from \cite{vqls} with tests on simulators.
In this work, we conduct a thorough study of the applicability of variational hybrid methods
to PDEs by performing extensive tests with VQLS for
the Poisson equation on both simulators with statistical finite sampling (shot) noise and real quantum hardware using superconducting qubits \footnote{IBM Quantum. \url{https://quantum-computing.ibm.com/}, 2022}.

Our results indicate that hybrid solvers for PDEs are not a promising route
for achieving quantum advantage in the short to near term.
It is well known that,
when increasing system size, a) PDEs require
preconditioning and b) variational algorithms suffer from
barren plateus. Neither of these issues are present for the system sizes
considered in this work.
Nonetheless, VQLS struggles to converge (fast) even for
small system sizes -- both on quantum hardware and
simulators with shot noise --
in the absence of a), b), data encoding and readout
issues.
Moreover, the competition -- namely classical PDE solvers -- can
typically achieve precision that seems beyond the reach of quantum algorithms relying on
finite sampling.

A common argument \emph{for} the potential of quantum advantage
is the scalability with respect to the number of spatial dimensions of the PDE
-- ignoring data en- and decoding, there is no apparent
``curse of dimensionality'' in VQLS applied to, e.g., the Poisson equation
\footnote{Nonetheless, other issues may render VQLS unusable, e.g., high number of variational
parameters or exponentially worse barren plateus. Moreover, the convergence rate of the estimation error
w.r.t.\ the number of shots is at best square root, i.e., analogous to classical
Monte Carlo methods.}.
However, for interesting system sizes of $n\geq 20$ qubits and the resulting increase
in noise,
we believe other classical methods for high-dimensional PDEs
\cite{Sirignano18, Han18, Bachmayr16}
are more promising than VQAs.
While there still may be some benefit in using NISQ devices for, e.g.,
material simulation or CFD, we believe it does not lie within the direct resolution of
PDEs via an amplitude encoding of the solution (see Section \ref{sec:vqls}).

Finally, we mention that in this work we considered several but not all error
mitigation techniques. Particularly noteworthy is probabilistic error cancellation (PEC)
\cite{Temme17, Berg22}
that was recently added to IBM's runtime service.
PEC attempts to produce unbiased estimates of expectation values
by fitting a (sparse) Pauli noise model to the physical noise
on the quantum device
and implementing the inverse of said noise channels by sampling randomized Pauli twirled circuits.
The success of this denoising method depends on several non-trivial assumptions about
the quality of the noise model and involves an exponential sampling overhead.
It is a potentially interesting research question
whether PEC would benefit the estimation of observables described in this work
and if it would scale to large system sizes.

The remainder of the paper is organized as follows.
In Section \ref{sec:poisson}, we introduce the Poisson equation and the different ways to
estimate the Poisson operator on a gate based quantum computer.
In Section \ref{sec:vqls}, we go over some basics of VQLS and the different
cost functions used for optimization.
In Section \ref{sec:results}, we test individual components of VQLS
for the Poisson equation -- both on simulators and quantum hardware -- and
conclude with the overall performance of the VQLS optimization in the
presence of noise.


\section{Poisson Equation}\label{sec:poisson}
The Poisson equation with Dirichlet boundary conditions is defined as
\begin{alignat*}{2}
	-\nabla^2u(x) &= f(x)&&\quad\text{for }x\in\Omega,\\
	u(x) &= 0&&\quad\text{for }x\in\partial\Omega,
\end{alignat*}
for an open domain $\Omega\subset\R^d$ with Lipschitz boundary
$\partial\Omega$ and real-valued $u:\bar\Omega\rightarrow\R$ and
$f:\Omega\rightarrow\R$.
The weak (variational) form of this equation reads:
find $u\in H^1_0(\Omega)$ that satisfies
\begin{align*}
	a(u,v)&:=\int_\Omega \nabla u(x)\cdot \nabla v(x)\diff x\\
	&=\int_\Omega f(x)v(x)\diff x=:f(v),
\end{align*}
for all $v\in H_0^1(\Omega)$,
where $H_0^1(\Omega)$ is the space of all weakly differentiable
$L^2$-functions that are zero on $\partial\Omega$ in the trace sense.

In the finite element method (FEM), one discretizes the Poisson equation
by choosing a set of basis (\emph{test} and \emph{trial}) functions
\begin{equation*}
	V_h:=\{\varphi_k:\;k=0,\ldots,N-1\}\subset H_0^1(\Omega),
\end{equation*}
and, consequently, the \emph{discrete} Poisson equation reads:
find $\bs u=(u_k)_{k=0}^{N-1}$ such that
$u_h:=\sum_{k=0}^{N-1}u_k\varphi_k$ and
\begin{equation*}
	a(u_h,v_h)=f(v_h),\quad\text{for all }v_h\in V_h,
\end{equation*}
or, equivalently, solve
\begin{align}\label{eq:lse}
	\bs A\bs u &= \bs f,\\
	\bs A&:=(a(\varphi_j,\varphi_i))_{i,j=0}^{N-1},\notag\\
	\bs f&:=(f(\varphi_i))_{i=0}^{N-1}.\notag
\end{align}

In this work, we test the simplest case $d=1$, $\Omega=(0,1)$ and piecewise linear
FEM. Therefore, $\bs A$ is a tridiagonal matrix with $2$'s on the main diagonal
and $-1$'s on the off-diagonals
\begin{equation}\label{eq:matrix}
\bs A=\frac{1}{h}
\begin{pmatrix}
	2  & -1     & 0      & \dots & \\
	-1 & 2      & -1     & \dots & \\
	   &        & \ddots &       & \\
	   &        &        &     & -1\\ 
	   &        &        & -1    & 2 
\end{pmatrix},
\end{equation}
where $h=\frac{1}{N+1}$ denotes the mesh size (distance between discretization nodes).
Alternatively, one can re-scale the domain $\Omega$ to $(0, N+1)$, in which case
$h=1$.

For the right-hand-side (RHS) $\bs f$, we consider the two border cases where $\bs f$
is constant and where $\bs f$ contains a discontinuous jump.
The constant case leads to a smooth solution $u$ whereas the discontinuous case leads to
a singularity (in the derivatives) of $u$. While this significantly impacts the performance
of classical methods (or requires special techniques such as, e.g., adaptive or hp-methods),
the discontinuity has no substantial effect on the training in Section \ref{sec:results}.
Both cases can be approximated as quantum states on $n$ qubits
with the following unitaries
\begin{align}
	\ket{\bs f_C} &:= H^{\otimes n}\ket{0}_n,\label{rhs:hn}\tag{Hn}\\
	\ket{\bs f_D} &:= H^{\otimes n-1}\otimes X\ket{0}_n,\label{rhs:hnx}\tag{HnX}
\end{align}
where $H$ and $X$ are the Hadamard and Pauli-$X$ gates, respectively.

We cast the LSE problem from \eqref{eq:lse} to a problem of determining a
quantum state via an amplitude encoding as follows.
Find an $n$-qubit
quantum state $\ket{\bs u}=\frac{1}{\|\bs u\|}\sum_{k=0}^{2^n-1}u_k\ket{k}$
that satisfies
\begin{equation}\label{eq:qlse}
	(\|\bs u\|)\bs A\ket{\bs u}=(\|\bs f\|)\ket{\bs f}.
\end{equation}
We re-scale such that $h\|\bs f\|=1$ for simplicity and since this does not
affect conclusions about the overall performance of VQLS.

Within a variational solver, one uses a parametrized ansatz $\ket{\psi(\theta)}$
and optimizes over the parameters $\theta=(\theta_1,\ldots,\theta_p)$ to approximate
the normalized solution $\ket{\bs u}\approx\ket{\psi(\theta^*)}$.
To this end, we must estimate terms such as $\braket{\psi(\theta)|\bs A|\psi(\theta}$
on a quantum computer.
There are several ways of estimating the observable $\bs A$: details
of all decompositions used in this work are provided in
Fig.~\ref{fig:IX} -- \ref{fig:Liu21Grouped}.

The simplest option is decomposing $\bs A$ into Pauli strings.
This requires $\mc O(2^n)$ terms and, consequently, any
potential quantum advantage is lost.
Another option was presented in \cite{Liu21}, where $\bs A$ is decomposed into
$\mc O(n)$ simple operators, see Fig.~\ref{fig:Liu21}. Note that, however,
the decomposition terms in Liu21 commute and can be thus grouped together into
$\mc O(1)$ terms, see Fig.~\ref{fig:Liu21Grouped}.
Finally, in \cite{Sato21}, yet another decomposition method was proposed with
$\mc O(1)$ terms.

While all of these decompositions are mathematically equivalent,
they have a trade-off between number of circuits to run vs. entangling gates
required for each circuit. The matrix $\bs A$ from \eqref{eq:matrix}
can be decomposed as
\begin{equation}\label{eq:Adecomp}
	\bs A = 2I^{\otimes n} - I^{\otimes n-1}\otimes X + R. 
\end{equation}
The expectation of the first term is constantly equal to 2. The second term requires
adding only one Hadamard gate on the least significant
qubit \footnote{By convention, the measurements on a quantum device are performed in the
$Z$-basis. Therefore, we require an $H$-gate to rotate into the $X$-basis.}, i.e., here the accuracy will mostly
depend on the gates required for the ansatz $\ket{\psi(\theta)}$.
The remaining term $R$ (off-diagonal $-1$'s on odd positions) requires either
exponentially many Pauli strings or one highly entangling circuit, see also
Fig.~\ref{fig:Sato21}, \ref{fig:Liu21Grouped}.
On one hand, using highly entangling unitaries, the estimation of the
expectation value of $R$ is inaccurate on today's quantum computers. On the other hand,
as we will see in Section \ref{sec:operator},
the expectation value of $\bs A$ is dominated by the constant part.
This is a good example of an abstract mathematical problem that is
deceivingly simple to solve classically,
but is rather complicated to run on a modern day quantum computer.


\section{VQLS Cost Functions}\label{sec:vqls}
VQLS was first introduced in \cite{vqls}. It is an extension of the variational
quantum eigensolver \cite{vqe}, where one uses a parametrized ansatz
function $\ket{\psi(\theta)}$ and minimizes the expected
energy of some Hamiltonian. The different types of ansaetze used in this work
are detailed in Fig.~\ref{fig:ansaetze}.

For linear systems, the authors in \cite{vqls} propose
Hamiltonians for which the ground state corresponds to the normalized solution
state from \eqref{eq:qlse}. 
One can also minimize more general cost functions that do not directly correspond
to Hamiltonian minimization, e.g., as for the cost function in \cite{Sato21}.
In this work, we will use one of the following four cost functions
\begin{align}\label{eq:costfs}
	C_{\mathrm N}(\theta)&:=-\frac{1}{2}\frac{(\Re\braket{\bs f|\psi(\theta)})^2}{\braket{\psi(\theta)|\bs A|\psi(\theta)}},\\
	C_{\mathrm{NN}}(\theta,s)&:=\frac{1}{2}s^4\braket{\psi(\theta)|\bs A|\psi(\theta)}
	-s^2\Re\braket{\bs f|\psi(\theta)},\notag\\
	C_{\mathrm G}(\theta)&:=\frac{\braket{\psi(\theta)|H_G|\psi(\theta)}}{\braket{\psi(\theta)|\bs A^\dag\bs A|\psi(\theta)}},\notag\\
	C_{\mathrm L}(\theta)&:=\frac{\braket{\psi(\theta)|H_L|\psi(\theta)}}{\braket{\psi(\theta)|\bs A^\dag\bs A|\psi(\theta)}},\notag
\end{align}
where, for $\ket{\bs f}=U_f\ket{0}$,
\begin{align*}
	H_G&:=\bs A^\dag U_f\left(I^{\otimes n}-\ket{0}\bra{0}\right)U_f^\dag\bs A,\\
	H_L&:=\bs A^\dag U_f\left(I^{\otimes n}-\frac{1}{n}\sum_{j=0}^{n-1}\ket{0_j}\bra{0_j}\otimes I_{\bar j}\right)U_f^\dag\bs A,
\end{align*}
with $\ket{0_j}\bra{0_j}$ being the projection onto the $j$-th qubit only and
$I_{\bar j}$ the identity on all except the $j$-th qubit.
The operational meaning, pros and cons of different cost functions have been previously discussed in
other works \cite{vqls, CerezoCost21, Cerezo22}.
The cost functions $C_{\mathrm G}$ and $C_{\mathrm L}$ determine the solution by minimizing
the residual $\bs f-\bs A\ket{\psi(\theta)}$ (disregarding re-scalings), while
$C_{\mathrm{N}}$ and $C_{\mathrm{NN}}$ minimize the Dirichlet energy of the solution candidate. 

Cost functions $C_{\mathrm G}$ and $C_{\mathrm L}$ were introduced in \cite{vqls}.
$C_{\mathrm N}$ was introduced in \cite{Sato21} and corresponds to the Dirichlet energy
for the normalized solution,
while $C_{\mathrm{NN}}$ is its non-normalized version, i.e.,
$s^2$ stands for the norm of $\bs u$.
Although the global minimum of all four cost functions corresponds to the normalized solution
of \eqref{eq:qlse}, the estimation accuracy and convergence of optimizers varies.
Moreover, we note that the cost functions $C_{\mathrm N}$ and $C_{\mathrm{NN}}$
require fewer and simpler circuits than $C_{\mathrm G}$ and $C_{\mathrm L}$.


\section{Results}\label{sec:results}
A VQLS setup consists of choosing the type of ansatz, the RHS, the number of qubits
and layers of the ansatz, the cost function, the classical optimization method
and the estimation backend.
In Section \ref{sec:nonoise}, we probe the expressivity and trainability of different
ansaetze and cost functions by testing on noiseless state-vector simulators, i.e.,
in the absence of both hardware noise and statistical finite sampling (shot) noise.
In Section \ref{sec:ansatz}, we test the fidelity of different ansaetze both on simulators
with shot noise and on different IBM backends.
In Section \ref{sec:innerp}, we test the accuracy of estimating inner products.
In Section \ref{sec:operator}, we test the accuracy of estimating the Poisson operator
expectation.
In Section \ref{sec:costf}, we test the accuracy of estimating different cost functions.
In Section \ref{sec:grad}, we test the cosine similarity of estimating gradients.
Finally, in Section \ref{sec:noise}, we test the overall performance of several optimizers
with different cost functions.
The tests in Sections \ref{sec:ansatz} -- \ref{sec:grad} were performed on a sample
of randomly generated ansatz parameters, where we compared, e.g., different error percentiles.

The solution fidelity is defined as
\begin{equation*}
	F(\theta):=|\braket{\psi(\theta)|\bs u}|^2,
\end{equation*}
where $\ket{\bs u}$ is the exact normalized solution.
We always compute the solution fidelity numerically exact, i.e., whether $\theta$ was
optimized on simulators or a real quantum device,
the above fidelity is computed using the exact vector for $\ket{\bs u}$
and the exact vector for $\ket{\psi(\theta)}$.

\subsection{Training Without Noise}\label{sec:nonoise}
In this section, we estimate all quantities on a state-vector simulator, i.e.,
numerically exact and choose the BFGS optimizer \cite{Broyden70, Fletcher70, Goldfarb70, Shanno70}.
It was observed both in previous work \cite{Jarman21} and our own tests that
BFGS performs best on small variational problems in the absence of noise.
We run each VQLS setup for 15 randomly generated initial values and select
the best run based on the smallest observed cost function value.
We use the same set of random initial values across all tests.

\begin{figure*}[t]
	\centering
	\begin{subfigure}{0.9\columnwidth}
		\includegraphics{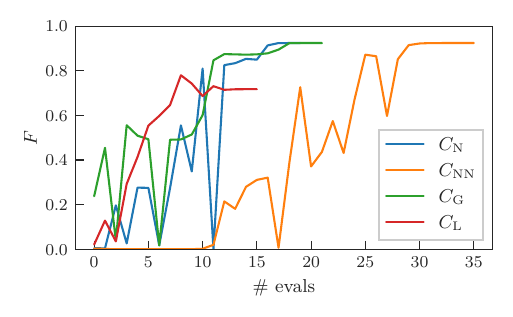}
		\caption{}
	\end{subfigure}\hfill
	\begin{subfigure}{0.9\columnwidth}
		\includegraphics{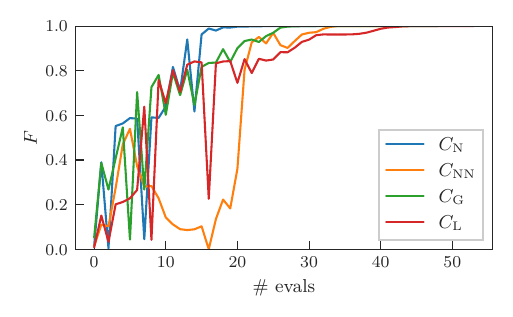}
		\caption{}
	\end{subfigure}
		\begin{subfigure}{0.9\columnwidth}
		\includegraphics{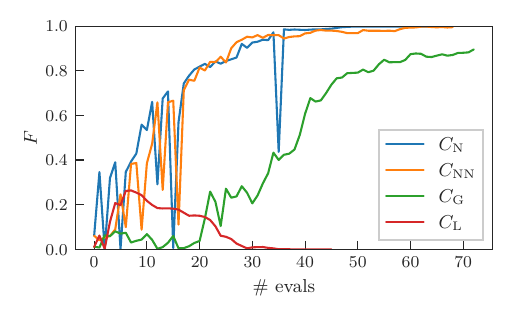}
		\caption{}
	\end{subfigure}\hfill
	\begin{subfigure}{0.9\columnwidth}
		\includegraphics{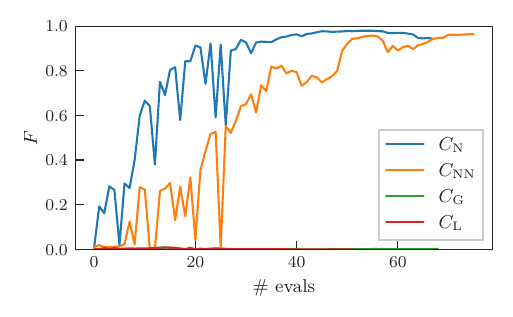}
		\caption{}
	\end{subfigure}
	\begin{subfigure}{0.9\columnwidth}
		\includegraphics{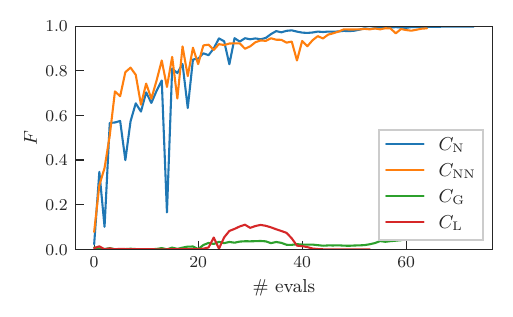}
		\caption{}
	\end{subfigure}\hfill
	\begin{subfigure}{0.9\columnwidth}
		\includegraphics{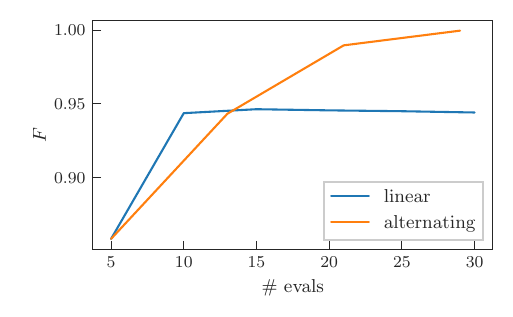}
		\caption{}
	\end{subfigure}
	\caption{Solution fidelity vs.\ number of cost function evaluations.
	Note that we display \emph{all} cost evaluations, i.e., not only
	cost for current (accepted) iterate $\theta_k$.
	RHS \eqref{rhs:hnx}, BFGS optimizer, state vector simulator (no noise).
	(a) Ansatz linear alternating $R_Y$-$CZ$
	(see Fig.~\ref{fig:ansaetze}), $n=3$, $l=0$.
	(b) Ansatz linear alternating $R_Y$-$CZ$, $n=3$, $l=3$.
	(c) Ansatz linear alternating $R_Y$-$CZ$, $n=5$, $l=3$.
	(d) Ansatz linear alternating $R_Y$-$CZ$, $n=8$, $l=3$.
	(e) Ansatz linear alternating $R_Y$-$CZ$ ansatz, RHS \eqref{rhs:hn},
	$n=5$, $l=3$.
	(f) Final solution fidelity vs. number of ansatz parameters,
	RHS \eqref{rhs:hnx}, $n=5$, RY-CZ ansatz, linear vs linear alternating.}
  \label{fig:vqa_exact}
\end{figure*}

In Fig.~\ref{fig:vqa_exact}, we display the solution fidelity for the minimization based
on the four different cost functions from \eqref{eq:costfs},
for a varying number of qubits and ansatz layers.
The cost function $C_{\mathrm N}$ leads to the fastest convergence.
The non-normalized version $C_{\mathrm{NN}}$ displays similar but slightly worse behavior.
Our main reason for considering $C_{\mathrm{NN}}$ is a better behavior in the presence of
noise, as we will see in the following sections.
The global and local cost functions, $C_{\mathrm G}$ and $C_{\mathrm L}$,
fail to converge at a relatively small number of qubits and are, in addition, more costly
to estimate than $C_{\mathrm N}$ and $C_{\mathrm{NN}}$.

The choice of the RHS does not pose a problem for the $C_{\mathrm N}$ and $C_{\mathrm{NN}}$ cost
functions. For $C_{\mathrm G}$ and $C_{\mathrm L}$, the barren plateaus worsen for
the RHS \eqref{rhs:hn}. This may seem counter-intuitive from the perspective of classical
PDE methods, where smoother data is generally associated with better convergence.
For VQLS, however, particular combinations of ansaetze and RHSs can lead
to very flat optimization landscapes or even constant cost functions.
The choice of alternating vs.\ non-alternating entanglement pattern from Fig.~\ref{fig:ansaetze} has
a significant impact on the final solution fidelity.

\subsection{Ansatz Fidelity}\label{sec:ansatz}
Having computed the optimal parameters, how accurately can we sample
the corresponding distribution from the quantum device?
In this section, we test the fidelity of different ansaetze across multiple backends.
As a benchmark, we will use the sampling error of a simulator with
shot noise.
For the error metric, we use the Hellinger fidelity of two distributions $p$ and $q$
\begin{equation*}
    F(p, q):=(\sum_i\sqrt{p_iq_i})^2=|\braket{\Tilde{\psi}(\theta)|\psi(\theta)}|^2,
\end{equation*}
where $p_i=|\braket{i|\psi(\theta)}|^2$,
$q_i=|\braket{i|\Tilde{\psi}(\theta)}|^2$,
$\ket{\psi(\theta)}$ is the true ansatz state
and $\ket{\Tilde{\psi}(\theta)}$ is the ansatz state prepared on a real quantum device.
More precisely, we have access only to an estimate $\hat{q_i}$ of the true probability
$q_i$.
The different ansatz architectures are described in Fig.~\ref{fig:ansaetze}.

\begin{figure*}[t]
	\centering
	\begin{subfigure}{0.9\columnwidth}
		\includegraphics{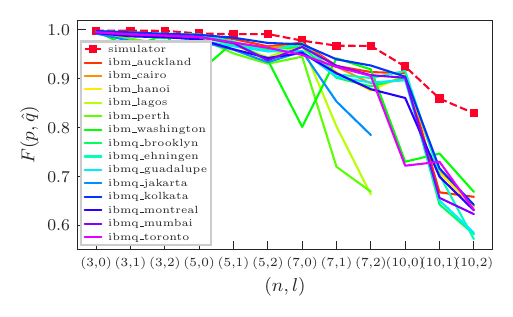}
		\caption{}
		\label{fig:cfbackends}
	\end{subfigure}\hfill
	\begin{subfigure}{0.9\columnwidth}
		\includegraphics{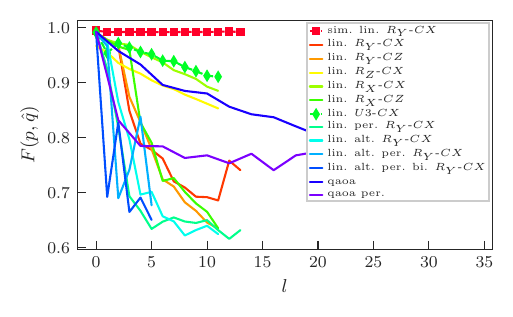}
		\caption{}
		\label{fig:cfarchts}
	\end{subfigure}
	\begin{subfigure}{0.9\columnwidth}
		\includegraphics{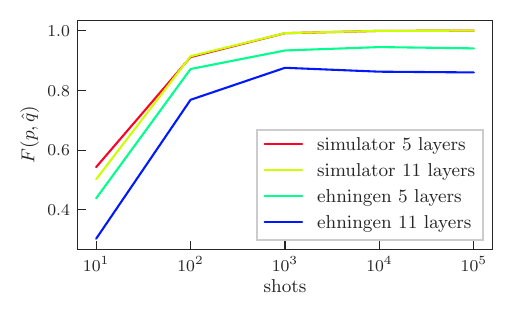}
		\caption{}
		\label{fig:ansatz_shots}
	\end{subfigure}\hfill
	\begin{subfigure}{0.9\columnwidth}
		\includegraphics{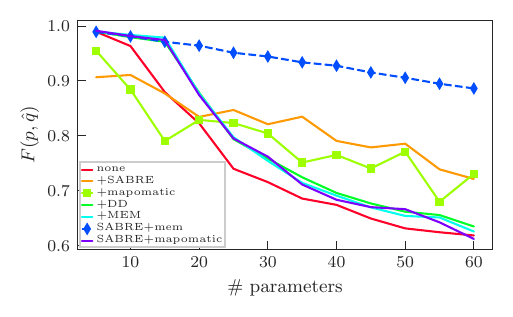}
		\caption{}
		\label{fig:ansatz_transpile}
	\end{subfigure}
	\caption{Tests with ansatz fidelity on real backends.
	Median fidelity out of 50 random parameter samples.
	For (a) -- (c) only optimized transpilation was applied.
	(a) Medial fidelity vs.\ (qubits $=n$, layers $=l$),
	number of parameters $=n\cdot (l+1)$, on different IBM backends.
	1000 shots.
	(b) Fidelity vs.\ number of parameters for different ansatz architectures with
	5 qubits on
	\texttt{ibmq\_ehningen}. 1000 shots.
	See Fig.~\ref{fig:ansaetze} for details on ansatz architectures.
	(c) Fidelity vs.\ number of shots for the linear $R_Y$-$CZ$ ansatz
	with 5 qubits.
	(d) Fidelity vs.\ number of parameters for different transpilation and
	error mitigation heuristics, linear $R_Y$-$CZ$ ansatz with 5 qubits.}
\end{figure*}

In Fig.~\ref{fig:cfbackends}, we see that the median fidelity is quite close to
the noiseless quantum simulator up to 10 qubits. Adding entangling layers at 10 qubits
reduces the fidelity significantly.
Comparing the different backends, the performance is similar,
with dips in fidelity
observed on either newer backends or on small backends where we exhausted the maximum
number of qubits.

In Fig.~\ref{fig:cfarchts}, we compare different ansatz architectures.
Some results are to be expected, e.g., adding periodic entanglement
(CNOTs on first and last qubit) leads to more CNOTs post
transpilation -- since the topology of IBM devices does not natively
accommodate such entanglement (see Fig.~\ref{fig:ibm_layouts}) -- and, hence, lowers fidelity.
Other results are, however, less intuitive.
For example, linear $R_X$-$CX$ performs slightly better than linear
linear $R_Z$-$CX$, even though $R_Z$ rotations are native to IBM hardware.
More notably, linear $U3$-$CX$ (all 3 rotations) performs best despite performing
more rotations. The number of parameters is tripled for
$U3$, i.e., for $p=180$, $U3$-$CX$ has as many entangling
layers as $R_Y$-$CZ$ for $p=60$.

In Fig.~\ref{fig:ansatz_shots},
we consider the fidelity behavior when increasing the number of shots
(measurement samples). The convergence on a noiseless quantum simulator
and \texttt{ibmq\_ehningen} are similar
with a constant offset due to hardware noise.
For the simulator, the number of entangling layers does not affect the fidelity, as expected,
whereas for the real backend adding layers accumulates hardware error. 

Finally, when executing circuits on a real backend, one has various transpilation
and error mitigation options. For comparison, we select those options that we consider
to be relatively generic and easy to run by most users \footnote{Recently, IBM extended their software tools by adding
runtime primitives with streamlined error mitigation.
Some noteworthy error mitigation techniques not considered in our tests are
zero noise extrapolation and probabilistic error cancellation \cite{Temme17, Berg22}.}.
For transpilation, we compare no optimization with
\texttt{Qiskit}'s swap-optimized transpilation based on the SABRE heuristic \cite{sabre}.
Due to the stochastic nature of this heuristic,
we transpile 20 times and select the circuit minimizing a weighted average
of depth and number of CNOTs, with number of CNOTs having twice the weight.
For error mitigation (EM), we test either no EM, matrix-free measurement EM (MEM) \cite{m3mem},
simple $XX$ dynamic decoupling (DD) with as-late-as-possible instruction scheduling
or noise aware transpilation (the \texttt{mapomatic} tool from \cite{mapomatic}).
The results are summarized in Fig.~\ref{fig:ansatz_transpile}.

Optimized transpilation and measurement error mitigation
consistently improve fidelity.
When an algorithm does not change the topology of input circuits,
transpilation has to be performed only once and, thus, the overhead is negligible.
For MEM, one has to perform regular calibrations on the target backend.
If integrated into the output pipeline,
the overhead for MEM can be negligible as well.

The results for dynamic decoupling are less clear.
From Fig.~\ref{fig:ansatz_transpile}, \ref{fig:innerp}
and other tests not illustrated here,
we observed that it can both improve and worsen the results.
We are not aware of a universal DD method that works on any circuit,
it is still an active are of research and a more
problem-tailored DD method might be required for consistent improvement
(see, e.g., \cite{Ravi22}).

The same conclusion applies to noise aware transpilation from \cite{mapomatic}.
Here, a better choice of cost function than the default
for selecting less noisy qubits might improve results.
We do not investigate this further.

\subsection{Inner Products}\label{sec:innerp}
Estimating inner products is a basic numerical subroutine
required by any linear solver. In our case,
we estimate inner products of the form
$\Re\braket{\bs f|\psi(\theta)}$. This is typically done via the Hadamard
test as in \cite{vqls, Sato21}.
For the special case of the cost function $C_{\mathrm{N}}$ from
\eqref{eq:costfs} and an ansatz with only real-valued amplitudes,
we can estimate $|\braket{\bs f|\psi(\theta)}|^2$ by measuring the overlap,
which requires fewer controlled gates than the Hadamard test.
The overlap test measures all qubits as opposed to
one qubit for the Hadamard test.

We compare relative errors since the exact value of
$\Re\braket{\bs f|\psi(\theta)}$ is often quite small and thus small absolute
errors are not informative.
Let $\mc E(\theta):=\mc E(\Re\braket{\bs f|\psi(\theta)})$
denote the estimate of the inner
product computed by sampling on a simulator or a real quantum device.
The exact definition of the relative error depends on whether we
compare the Hadamard and overlap test in the same plot,
or only the Hadamard test.
For the former, the relative error is defined as
\begin{equation*}
    \epsilon^2_{\mathrm{rel}}(\theta):=\frac{\left| |\braket{\bs f|\psi(\theta}|^2-\mc E(\theta)^2\right|}{|\braket{\bs f|\psi(\theta}|^2},
\end{equation*}
and for the latter
\begin{equation*}
    \epsilon_{\mathrm{rel}}(\theta):=\frac{\left| \Re\braket{\bs f|\psi(\theta}|-\mc E(\theta)\right|}{|\Re\braket{\bs f|\psi(\theta}|}.
\end{equation*}
This is simply because, when using the overlap test, we only
have access to the squared quantity and thus a direct comparison
of $\mc E(\theta)$ with $\mc E(\theta)^2$ in the same plot is not meaningful.

\begin{figure*}[t]
	\centering
	\begin{subfigure}{0.9\columnwidth}
		\includegraphics{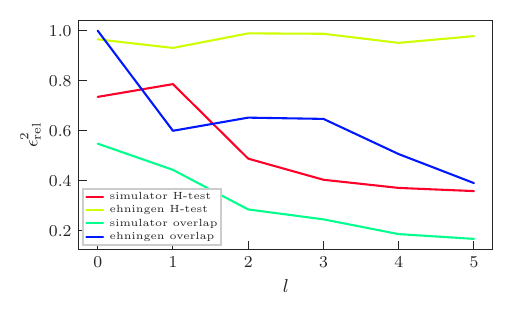}
		\caption{}
		\label{fig:innerpmethods}
	\end{subfigure}\hfill
	\begin{subfigure}{0.9\columnwidth}
		\includegraphics{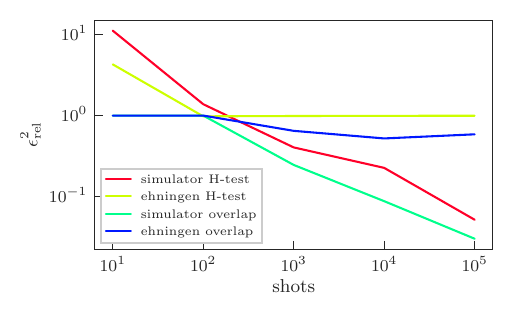}
		\caption{}
	\end{subfigure}
	\begin{subfigure}{0.9\columnwidth}
		\includegraphics{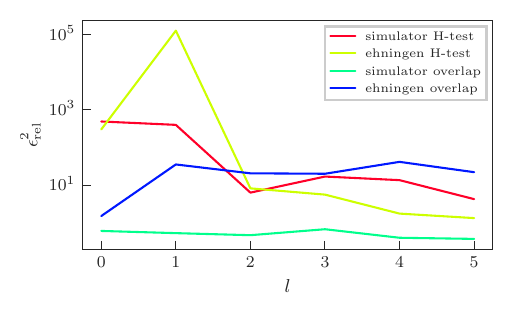}
		\caption{}
	\end{subfigure}\hfill
	\begin{subfigure}{0.9\columnwidth}
		\includegraphics{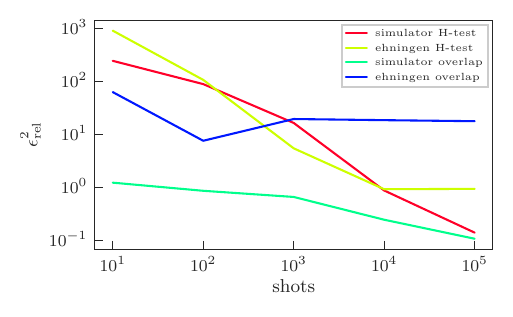}
		\caption{}
		\label{fig:innerpshotsmean}
	\end{subfigure}\hfill
	\begin{subfigure}{0.9\columnwidth}
		\includegraphics{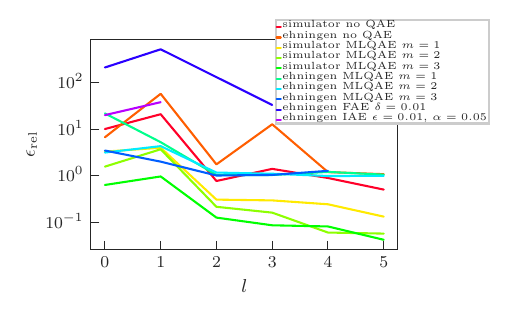}
		\caption{}
		\label{fig:innerpqae}
	\end{subfigure}\hfill
	\begin{subfigure}{0.9\columnwidth}
		\includegraphics{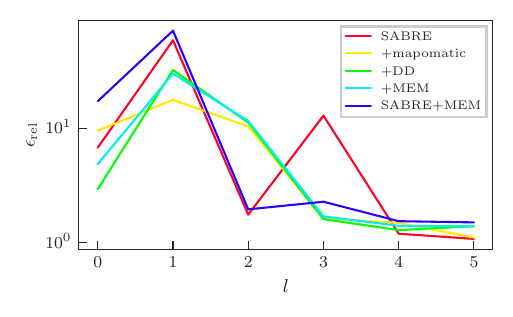}
		\caption{}
		\label{fig:innerptranspile}
	\end{subfigure}
	\caption{Relative error inner product estimation.
	50 parameter samples in all plots.
	(a) Median relative error vs.\ number of layers, $n=5$ qubits, linear $R_Y$-$CZ$ ansatz,
	RHS \eqref{rhs:hn}, 1000 shots.
	(b) Median relative error vs.\ number of shots, 5 qubits and 3 layers,
	linear $R_Y$-$CZ$ ansatz,
	RHS \eqref{rhs:hn}, 1000 shots.
	(c) Same as (a) but mean instead of median.
	(d) Same as (b) but mean instead of median.
	(e) Mean relative error vs.\ number of layers for the Hadamard test for different QAE methods,
	$R_Y$-$CZ$ ansatz, 5 qubits, RHS \eqref{rhs:hn}.
	(f) Mean relative error vs.\ number of layers for the Hadamard test for different error mitigation techniques,
	$R_Y$-$CZ$ ansatz, 5 qubits, RHS \eqref{rhs:hn}.
	}
\end{figure*}

In Fig.~\ref{fig:innerpmethods} -- \ref{fig:innerpshotsmean}, we compare
the different methods w.r.t.\ the number of layers and shots with references to
the exact value $|\braket{\bs f|\psi(\theta)}|^2$.
The conclusions are different depending on whether one considers only the median
or the mean. The overlap test performs slightly better if one ignores outliers,
and vice versa for the Hadamard test.
For the median, results do not improve significantly for more than 1000 shots,
for the mean, the threshold seems to be 10000 shots for the Hadamard test.

In Fig.~\ref{fig:innerpqae}, we consider different quantum amplitude estimation (QAE)
techniques for the Hadamard test.
Maximum likelihood quantum amplitude estimation (MLQAE) improves accuracy.
Note, however, that inner product estimation within VQLS is applied as a subroutine several times
per optimizer step. Therefore, on today's quantum computing hardware,
the additional overhead for adding QAE to each subroutine call is substantial for the overall
training, while the accuracy gain is potentially negligible.

Finally, in Fig.~\ref{fig:innerptranspile}, we consider adding different error mitigation
techniques for the Hadamard test.
Unlike in Section \ref{sec:ansatz}, the mean relative error is smallest when
applying all error mitigation techniques, while applying only optimized transpilation with
measurement error mitigation performs almost as well.

Although there is a lot of variability in errors depending on ansatz architecture, RHS,
number of layers and error mitigation techniques -- in all cases the relative error
for estimating the basic quantity  $\Re\braket{\bs f|\psi(\theta)}$ is very large,
even on simulators.
Furthermore, inner product estimation is a basic subroutine that would be required
to estimate functionals of the solution such as, e.g., taking the average of $\bs u$,
and this is a necessary step to gain any sort of quantum advantage for
PDE solvers \cite{hhl_fem}.
Based on our tests, even if the solution $\ket{\bs u}$ is accurately approximated
on a quantum device via $\ket{\psi(\theta)}$,
the necessary accuracy for estimating $\braket{r|\psi(\theta)}$ for some functional
$r$ and $n\geq 20$ qubits
translate into a very large number of shots together with high requirements
for hardware fidelity and error mitigation that may be unrealistic in the near term.

\subsection{Operator Expectation}\label{sec:operator}
In this section, we test the accuracy of estimating
$\braket{\psi(\theta)|\bs A|\psi(\theta)}$. As discussed in Section
\ref{sec:poisson}, there are at least four different ways to estimate $\bs A$,
refer to Fig.~\ref{fig:IX} -- \ref{fig:Liu21Grouped} for the different decompositions.

\begin{figure*}[t]
	\centering
	\begin{subfigure}{0.9\columnwidth}
		\includegraphics{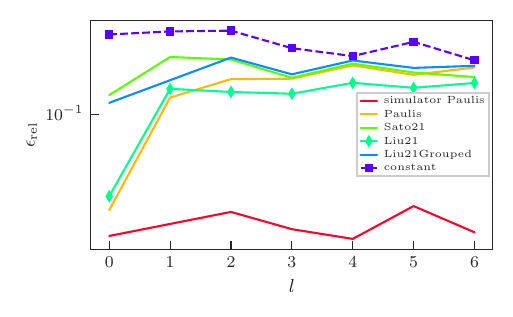}
		\caption{}
	\end{subfigure}\hfill
	\begin{subfigure}{0.9\columnwidth}
		\includegraphics{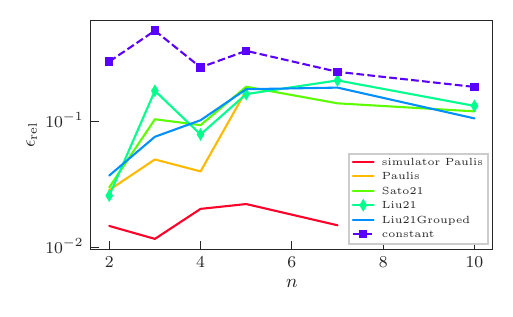}
		\caption{}
	\end{subfigure}\hfill
	\begin{subfigure}{0.9\columnwidth}
		\includegraphics{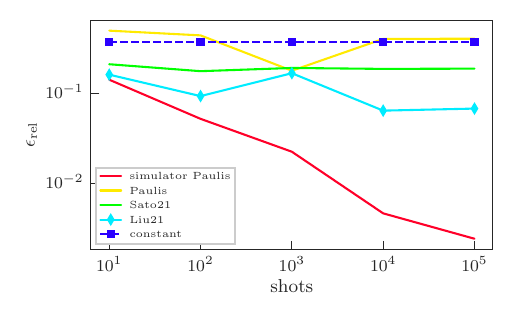}
		\caption{}
		\label{fig:op_shots}
	\end{subfigure}\hfill
	\begin{subfigure}{0.9\columnwidth}
		\includegraphics{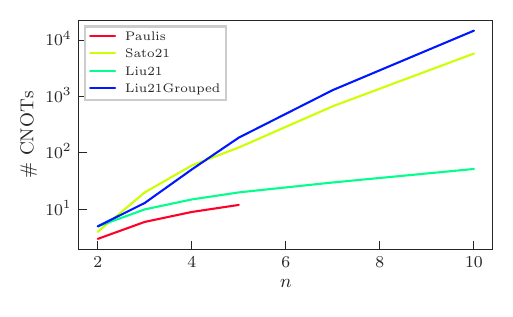}
		\caption{}
		\label{fig:op_cnots}
	\end{subfigure}
	\caption{50 parameter samples in (a) -- (c).
	(a) Mean relative error vs.\ number of layers,
	linear $R_Y$-$CZ$, 5 qubits, 1000 shots.
	(b) Mean relative error vs.\ number of qubits, linear $R_Y$-$CZ$, 3 layers, 1000 shots.
	(c) Mean relative error vs.\ number of shots, linear alternating $R_Y$-$CZ$.
	(d) Number of CNOTs vs.\ number of qubits, linear alternating $R_Y$-$CZ$,
	3 layers.}
	\label{fig:op}
\end{figure*}

In Fig.~\ref{fig:op}, we compare the relative errors defined as
\begin{equation*}
    \epsilon_{\mathrm{rel}}(\theta):=\frac{\left|\braket{\psi(\theta)|\bs A|\psi(\theta)}-\mc E(\theta)\right|}{|\braket{\psi(\theta)|\bs A|\psi(\theta)}|},
\end{equation*}
where we now abbreviate the estimate computed by sampling
on a simulator or a real quantum device as $\mc E(\theta):=\mc E(\braket{\psi(\theta)|\bs A|\psi(\theta)})$.
At first glance, this estimation seems to perform
much better than inner product estimation from Section \ref{sec:innerp}.
However, this is mainly due to the ``constant'' part of $\bs A$, see
\eqref{eq:Adecomp}, $\braket{\psi(\theta)|\bs A|\psi(\theta)}\approx 2$.

Fig.~\ref{fig:op_shots} shows that most of the error is hardware noise, e.g., for
$q=5$, $l=3$ it is not worth going beyond 100 -- 1000 shots.
Fig.~\ref{fig:op_cnots} shows the number of CNOTs required for the different decomposition methods.
Together with Fig.~\ref{fig:op_shots},
the Liu21 decomposition (see Fig.~\ref{fig:Liu21}) performs best.
We do not consider using the Pauli decomposition a viable option, since it does not scale.
Recall that the number of observables for Liu21 scales linearly with
the number of qubits as opposed to constant for Sato21 and Liu21Grouped.

\subsection{Cost Functions}\label{sec:costf}
In this section, we compare the accuracy of estimating $C_{\mathrm{N}}$ and $C_{\mathrm{NN}}$
from \eqref{eq:costfs}.
For inner product estimation, we use the Hadamard test, and, for operator expectation,
we use the Liu21 decomposition.
For a cost function $C(\theta)$ and its estimate $\hat{C}(\theta)$,
the absolute and relative errors are defined as
\begin{align*}
    \epsilon_{\mathrm{abs}}(\theta)&:=|C(\theta)-\hat{C}(\theta)|,\\
    \epsilon_{\mathrm{rel}}(\theta)&:=\epsilon_{\mathrm{abs}}(\theta)/|C(\theta)|.
\end{align*}

\begin{figure*}[t]
	\centering
	\begin{subfigure}{0.9\columnwidth}
		\includegraphics{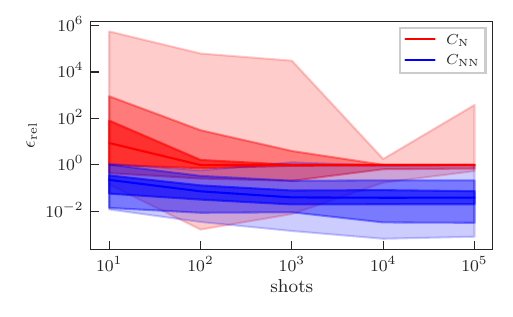}
		\caption{}
		\label{fig:cost_shots}
	\end{subfigure}\hfill
	\begin{subfigure}{0.9\columnwidth}
		\includegraphics{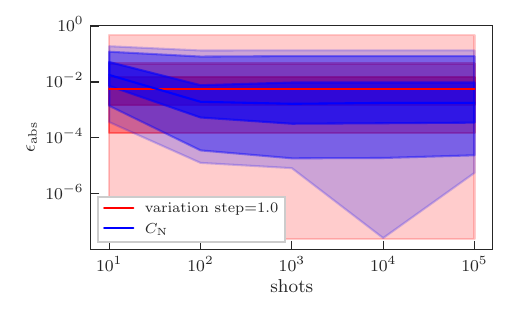}
		\caption{}
		\label{fig:cost_innerph1}
	\end{subfigure}\hfill
	\begin{subfigure}{0.9\columnwidth}
		\includegraphics{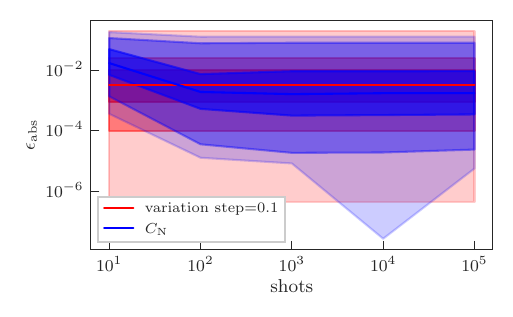}
		\caption{}
		\label{fig:cost_innerph01}
	\end{subfigure}\hfill
	\begin{subfigure}{0.9\columnwidth}
		\includegraphics{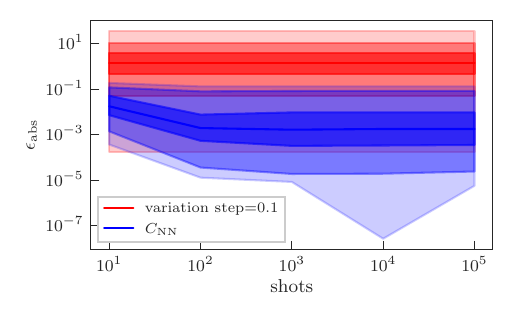}
		\caption{}
		\label{fig:cost_nonnormh01}
	\end{subfigure}
	\caption{Error percentiles 0, 5, 25, 50, 75, 95 and 100
	Linear alternating $R_Y$-$CZ$, 5 qubits, 3 layers.	
	(a) Relative error vs.\ number of shots.
	(b) -- (d) Absolute error vs.\ number of shots.}
\end{figure*}

In Fig.~\ref{fig:cost_shots},
we compare the accuracy w.r.t.\ the number of shots.
Unlike before, we show a more detailed comparison here,
including the $p=0, 5, 25, 50, 75, 95, 100$ error percentiles.
We see that the accuracy saturates at about $S=1000$ shots.
Moreover, estimation of $C_{\mathrm{NN}}$ is slightly more accurate.

Next, we test if the mean absolute error in cost function estimation is larger than the
mean variation of exact cost function values.
This should indicate if an optimizer can detect a descent direction in the presence of
hardware noise. We estimate the variation as follows. Generate a sample of $50$ random parameters
$\Omega:=\{\theta^1,\ldots,\theta^{50}\}$. Then, for each parameter in $\Omega$, sample
$100$ random parameters $\Delta(\theta):=\{\delta^1(\theta),\ldots,\delta^{100}(\theta)\}$,
where each of the components of $\delta^k(\theta)$ is between $-2\pi$ and $2\pi$.
Select a step size and compute the variations
\begin{equation*}
	var(\theta, \delta) := |C(\theta)-C(\theta+\mathrm{step}\cdot \delta)|,
\end{equation*}
for each $\delta$ in $\Delta(\theta)$ and each $\theta$ in $\Omega$.
In total, we thus have differences of cost function values sampled at 500 different points.

The results are summarized in Fig.~\ref{fig:cost_innerph1} -- \ref{fig:cost_nonnormh01}.
In both cases the estimation is quite noisy as
the absolute error in cost function value is within the range
of cost function variation. Thus, in general, it is unclear if an optimizer
can recognize a descent direction in such a noisy regime.

The results for $C_{\mathrm{NN}}$ are slightly better.
This is mostly due to the norm parameter $s$ in \eqref{eq:costfs}.
We thus expect that, for random initial values,
an optimizer would at first perform better for $C_{\mathrm{NN}}$.
However, as the optimization progresses and the value of $s$ is improved,
$C_{\mathrm{NN}}$ should encounter the same noise issues as $C_{\mathrm{N}}$.

\subsection{Gradients}\label{sec:grad}
\begin{figure*}[t]
	\centering
	\begin{subfigure}{0.9\columnwidth}
		\includegraphics{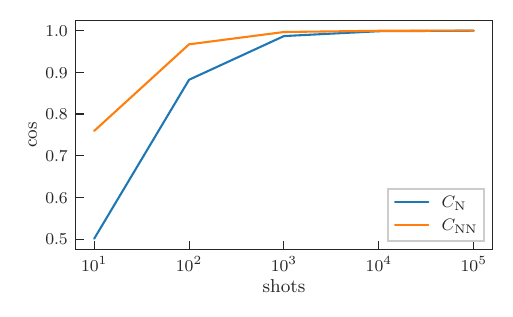}
		\caption{}
	\end{subfigure}\hfill
	\begin{subfigure}{0.9\columnwidth}
		\includegraphics{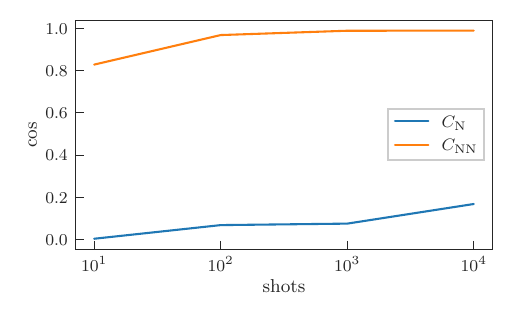}
		\caption{}
		\label{fig:grad_montreal}
	\end{subfigure}
	\caption{Cosine similarity vs.\ number of shots for different cost functions.
	We use the Hadamard test for inner product estimation and the Liu21
	decomposition for $\bs A$. Linear alternating $R_Y$-$CZ$ ansatz.
	(a) Median cosine similarity on a simulator for 50 random parameter samples,
	3 layers for the ansatz.
	(b) Median cosine similarity on \texttt{ibmq\_montreal} for 10 random parameter samples,
	1 layer for the ansatz.}
	\label{fig:grad}
\end{figure*}

In this section, we test the cosine similarity of estimated vs.\ exact
gradients for $C_{\mathrm{N}}$ and $C_{\mathrm{NN}}$.
The cosine similarity of two vectors $x$ and $y$ is defined as $\braket{x|y}/(\|x\|\|y\|)$.
The results are summarized in Fig.~\ref{fig:grad}.

Per optimizer iteration, often several calls of gradient estimation
are required. For each gradient call,
the number of circuits to run is $(C_1 + C_2p)S$,
where $S$ is the number of shots,
$p$ is the number of ansatz parameters and
$C_1$, and $C_2$ are constants depending on the cost function estimation method.
For, e.g., the Liu21 decomposition, $C_1=n+1$ and $C_2=2n+1$.
On today's quantum computers, this is a considerable overhead.
Thus, for the results in Fig.~\ref{fig:grad_montreal} on a real backend, we only use one ansatz layer and 10 parameter samples (in total $1490$ circuits to run $S$ times each).
As we can see, the gradient accuracy on \texttt{ibmq\_montreal} is very poor for
$C_{\mathrm N}$ and much better for $C_{\mathrm{NN}}$. For the latter,
we will see in the next section whether this is sufficient for an optimizer to find
a good solution.

\subsection{Training With Noise}\label{sec:noise}
In this section, our VQLS setup consists of a linear alternating
$R_Y$-$CZ$ ansatz with $n=5$ qubits and $l=3$ layers ($p=29$ parameters),
RHS \eqref{rhs:hnx}. We have seen in Section \ref{sec:nonoise} that in this
setting the ansatz is capable of representing the solution with near to
100\% fidelity.
We use the Hadamard test to estimate inner products for both $C_{\mathrm N}$
and $C_{\mathrm{NN}}$, and the Liu21 decomposition to estimate
$\braket{\psi(\theta)|\bs A|\psi(\theta)}$. We run the tests on a simulator
with shot noise and the \texttt{ibmq\_ehningen} backend.
We transpile all circuits as mentioned in Section \ref{sec:ansatz}.
For results on the quantum device, we present different combinations of
transpilation and measurement error mitigation options.
Note that we display the fidelity for \emph{all} parameters $\theta$ at which
the cost was evaluated, i.e., not only
for accepted iterates $\theta_k$.

There is a plethora of classical optimizers one could use, many are implemented in
\texttt{Qiskit} (most are wrappers for the \texttt{SciPy} package).
Based on the results of Section \ref{sec:nonoise},
other works \cite{Jarman21, Nakanishi20, Oliv22, Kandala17}
and our own tests, we only present here the BFGS \cite{Broyden70, Fletcher70, Goldfarb70, Shanno70},
SPSA \cite{spsa}, Powell \cite{Powell} and NFT \cite{Nakanishi20} optimizers.
Other optimizers do not seem to perform better in the presence of hardware noise.
Another possible noise-robust candidate -- that we did not test here -- is the
Bayesian optimizer from \cite{Iannelli22}.

\begin{figure*}[t]
	\centering
	\begin{subfigure}{0.9\columnwidth}
		\includegraphics{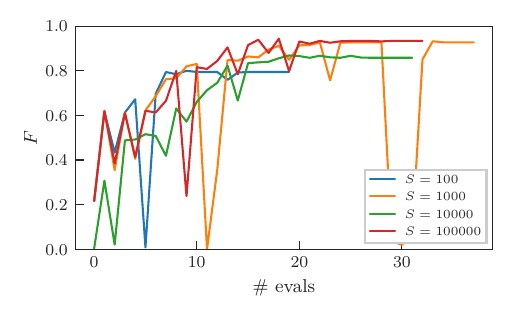}
		\caption{}
	\end{subfigure}\hfill
	\begin{subfigure}{0.9\columnwidth}
		\includegraphics{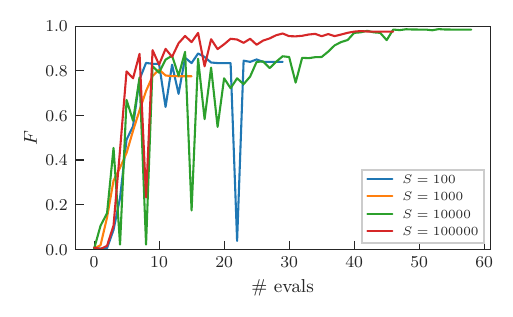}
		\caption{}
	\end{subfigure}\hfill
	\begin{subfigure}{0.9\columnwidth}
		\includegraphics{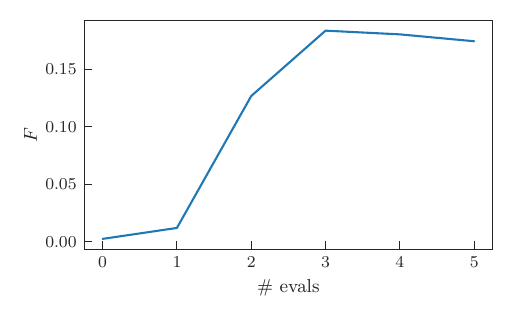}
		\caption{}
	\end{subfigure}
	\caption{VQLS training on simulators with shot noise and \texttt{ibmq\_ehningen},
	 BFGS optimizer with default hyperparameters,
	 linear alternating $R_Y$-$CZ$ ansatz, 5 qubits and 3 layers,
	 RHS \eqref{rhs:hnx}, Hadamard test for inner product estimation and
	 Liu21 for $\bs A$ decomposition.
	 (a) Fidelity vs.\ number of cost function evaluations (gradient evaluations not displayed)
	 on a simulator, training based on $C_{\mathrm{N}}$.
	 Best out of 15 initial values.
	 (b) Fidelity vs.\ number of cost function evaluations (gradient evaluations not displayed)
	 on a simulator, training based on $C_{\mathrm{NN}}$.
	 Best out of 15 initial values.
	 (c) Fidelity vs.\ number of cost function evaluations (gradient evaluations not displayed)
	 on \texttt{ibmq\_ehningen}, training based on $C_{\mathrm{NN}}$. Optimized noise-aware
	 transpilation with SABRE and \texttt{mapomatic}, see also
	 Section \ref{sec:ansatz}, MEM applied, 10000 shots per circuit.
	 Best out of 5 initial values.}
	\label{fig:vqa_bfgs}
\end{figure*}

\begin{figure*}[t]
	\centering
	\begin{subfigure}{0.9\columnwidth}
		\includegraphics{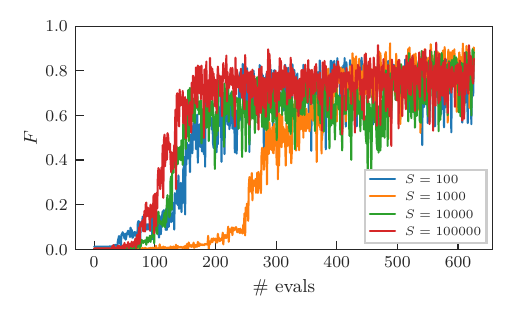}
        \caption{}
	\end{subfigure}\hfill
	\begin{subfigure}{0.9\columnwidth}
		\includegraphics{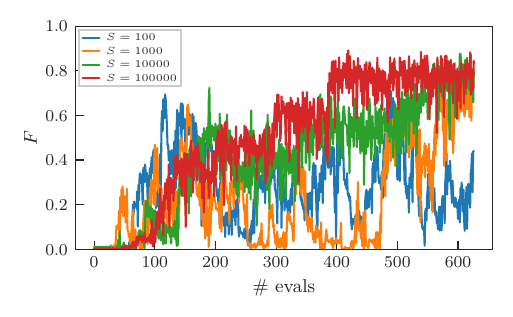}
		\caption{}
	\end{subfigure}\hfill
	\begin{subfigure}{0.9\columnwidth}
		\includegraphics{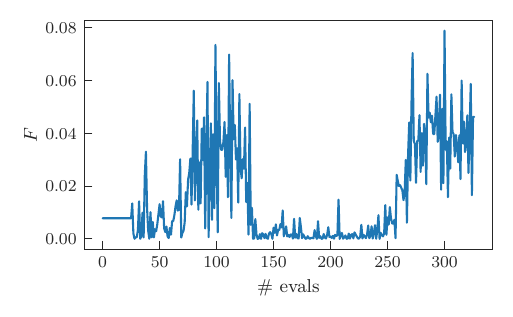}
		\caption{}
	\end{subfigure}
	\caption{VQLS training on simulators with shot noise and \texttt{ibmq\_ehningen},
	 SPSA optimizer with learning rate 1, perturbation 0.1, blocking and trust region set,
	 linear alternating $R_Y$-$CZ$ ansatz, 5 qubits and 3 layers,
	 RHS \eqref{rhs:hnx}, Hadamard test for inner product estimation and
	 Liu21 for $\bs A$ decomposition.
	 (a) Fidelity vs.\ number of cost function evaluations
	 on a simulator, training based on $C_{\mathrm{N}}$.
	 Best out of 15 initial values.
	 (b) Fidelity vs.\ number of cost function evaluations
	 on a simulator, training based on $C_{\mathrm{NN}}$.
	 Best out of 15 initial values.
	 (c)  Fidelity vs.\ number of cost function evaluations
	 on \texttt{ibmq\_ehningen}, training based on $C_{\mathrm{NN}}$. Optimized noise-aware
	 transpilation with SABRE, no MEM applied, 1000 shots per circuit.
	 Best out of 5 initial values.}
	\label{fig:vqa_spsa}
\end{figure*}

\begin{figure*}[t]
	\centering
	\begin{subfigure}{0.9\columnwidth}
		\includegraphics{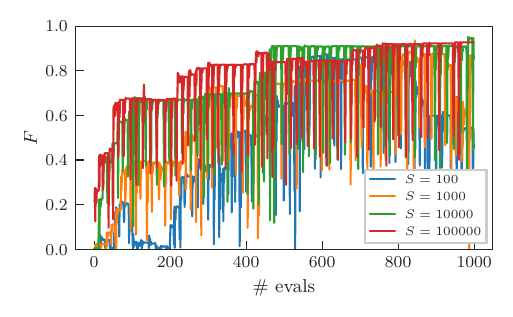}
		\caption{}
	\end{subfigure}\hfill
	\begin{subfigure}{0.9\columnwidth}
		\includegraphics{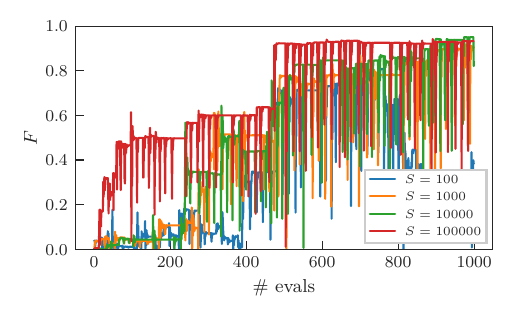}
		\caption{}
	\end{subfigure}\hfill
	\begin{subfigure}{0.9\columnwidth}
		\includegraphics{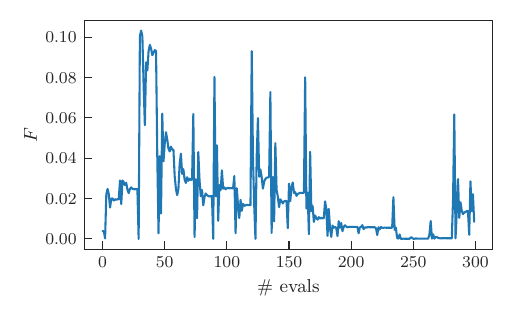}
		\caption{}
	\end{subfigure}
	\caption{VQLS training on simulators with shot noise and \texttt{ibmq\_ehningen},
	 POWELL optimizer with default hyperparameters,
	 linear alternating $R_Y$-$CZ$ ansatz, 5 qubits and 3 layers,
	 RHS \eqref{rhs:hnx}, Hadamard test for inner product estimation and
	 Liu21 for $\bs A$ decomposition.
	 (a) Fidelity vs.\ number of cost function evaluations
	 on a simulator, training based on $C_{\mathrm{N}}$.
	 Best out of 15 initial values.
	 (b) Fidelity vs.\ number of cost function evaluations
	 on a simulator, training based on $C_{\mathrm{NN}}$.
	 Best out of 15 initial values.
	 (c)  Fidelity vs.\ number of cost function evaluations
	 on \texttt{ibmq\_ehningen}, training based on $C_{\mathrm{NN}}$. Optimized noise-aware
	 transpilation with SABRE, no MEM applied, 10000 shots per circuit.
	 Best out of 5 initial values.}
	\label{fig:vqa_powell}
\end{figure*}

\begin{figure*}[t]
	\centering
	\begin{subfigure}{0.9\columnwidth}
		\includegraphics{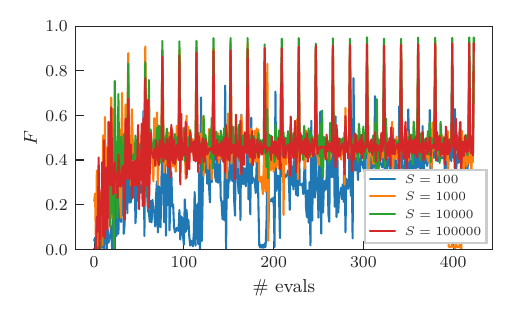}
		\caption{}
	\end{subfigure}\hfill
	\begin{subfigure}{0.9\columnwidth}
		\includegraphics{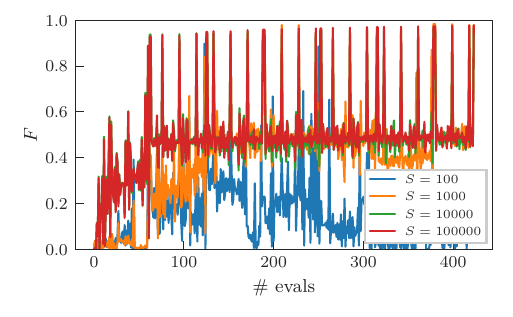}
		\caption{}
	\end{subfigure}\hfill
	\begin{subfigure}{0.9\columnwidth}
		\includegraphics{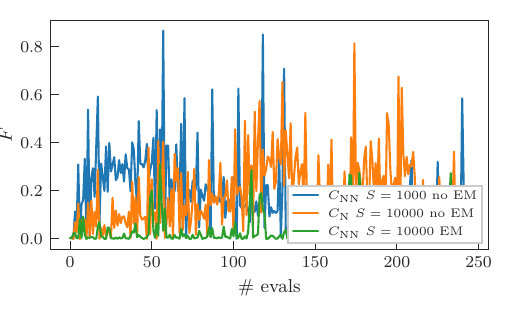}
		\caption{}
	\end{subfigure}
	\caption{VQLS training on simulators with shot noise and \texttt{ibmq\_ehningen},
	 NFT optimizer with reset interval 9,
	 linear alternating $R_Y$-$CZ$ ansatz, 5 qubits and 3 layers,
	 RHS \eqref{rhs:hnx}, Hadamard test for inner product estimation and
	 Liu21 for $\bs A$ decomposition.
	 (a) Fidelity vs.\ number of cost function evaluations
	 on a simulator, training based on $C_{\mathrm{N}}$.
	 Best out of 15 initial values.
	 (b) Fidelity vs.\ number of cost function evaluations
	 on a simulator, training based on $C_{\mathrm{NN}}$.
	 Best out of 15 initial values.
	 (c)  Fidelity vs.\ number of cost function evaluations
	 on \texttt{ibmq\_ehningen}. ``No EM'' means only optimized transpilation with
	 SABRE and ``EM'' means additionally noise-aware transpilation with \texttt{mapomatic}
	 and MEM.
	 Best out of 5 initial values.}
	\label{fig:vqa_nft}
\end{figure*}

In Fig.~\ref{fig:vqa_bfgs} -- \ref{fig:vqa_nft}, we summarize our findings.
BFGS performs significantly better for $C_{\mathrm{NN}}$,
achieving over $98\%$ fidelity for 10000 shots on a simulator.
On a quantum computer, for $C_{\mathrm{NN}}$, BFGS achieves roughly $17\%$ fidelity
but does not improve after that.
Moreover, the additional cost of evaluating gradients (see also
Section \ref{sec:grad}) is very time consuming on current quantum hardware:
for the linear alternating $R_Y$-$CZ$ ansatz on 5 qubits with 3 layers
and the Liu21 decomposition for $\bs A$,
one has to run 325 circuits $S$ times to estimate one gradient.
The few iterations as displayed in Fig.~\ref{fig:vqa_bfgs} require hours to execute
on \texttt{ibmq\_ehningen}.
Other gradient based methods implemented in \texttt{Qiskit}
failed to converge even with shot noise only
(for a moderate number of shots $S\leq 10^5$).

SPSA does not converge with the default choice of hyperparameters, i.e.,
using the calibration procedure from \cite{Kandala17}. Setting the learning rate to 1,
perturbation to 0.1, blocking and trust region to true,
we obtain the convergence displayed in Fig.~\ref{fig:vqa_spsa}.
We were not able to find a hyperparameter choice that works on a quantum computer.
With QN-SPSA \cite{qnspsa}, we obtained similar results.
SPSA convergence might be improved by an adaptive
hyperparameter selection as in, e.g., \cite{Sack22}, but we do not investigate this further.

Finally, in Fig.~\ref{fig:vqa_powell} and \ref{fig:vqa_nft}, we present the results for Powell and NFT, respectively.
The only tunable hyperparameter for NFT is the reset interval: a smaller interval generally leads
to smoother convergence but more cost function evaluations. In this example, we set the
reset interval of NFT to 9.
On simulators, NFT trained with $C_{\mathrm{NN}}$ performed best, achieving almost $98\%$ fidelity with only 1000 shots per circuit. A similar fidelity was obtained with BFGS trained
on $C_{\mathrm{NN}}$ with 10000 shots per circuit, at the expense of additionally estimating
gradients.
On \texttt{ibmq\_ehningen},
NFT trained on $C_{\mathrm{NN}}$ also performed best.
Although the maximum observed fidelity was over $80\%$, the fidelity corresponding to the best
observed cost value is only $10.89\%$ ($C_{\mathrm{N}}$ trained with no EM).
Similar observations about the noise robustness of NFT were made in
\cite{Nakanishi20, Oliv22}, however, overall, all trainings on quantum devices performed very poorly.

\begin{figure*}[t]
	\centering
	\includegraphics[width=0.67\columnwidth]{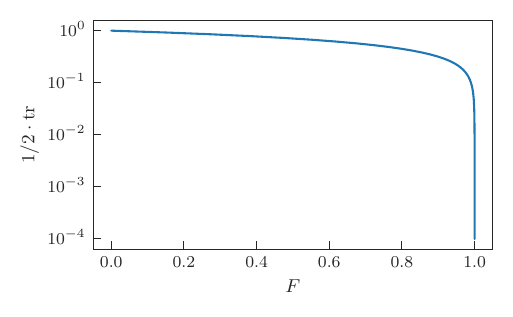}
	\caption{Trace distance ($y$-axis) vs.\ fidelity ($x$-axis).
	The last point on the $x$-axis is 0.99999999.}
	\label{fig:fidvstrace}
\end{figure*}

As discussed in the Introduction and Section \ref{sec:innerp},
we are ultimately interested in estimating inner products with linear functionals.
To that end, a more appropriate error metric would be the trace distance.
Since we are dealing with pure quantum states, there is a simple relationship between
the two
\begin{align*}
	&\frac{1}{2}\mathrm{tr}\left(
	\bigg|\;\ket{\bs u}\bra{\bs u} - \ket{\psi(\theta)}\bra{\psi(\theta)}\;\bigg|\right)
	\\
	&=\sqrt{1-F(\ket{\bs u}, \ket{\psi(\theta)})},
\end{align*}
where $F(\ket{\bs u}, \ket{\psi(\theta)})$ is the fidelity between the two pure states.
Consequently, a fidelity of $10.89\%$ translates into a trace distance of
$0.944$, which is a very large error.
For convenience, we plot the relationship between fidelity and trace distance
in Fig.~\ref{fig:fidvstrace}. For a reasonable trace error of $\leq 10^{-1}$, one would
require a solution fidelity of at least $99\%$.
This does not include the error of estimating
the inner product $\braket{r|\psi(\theta)}$ itself.

\section{Conclusion}
We conducted a thorough investigation of the feasibility of applying
modern-day quantum computers to partial differential equations.
Specifically, we have extensively tested the variational quantum linear solver for the simple
Poisson problem using IBM's superconducting quantum devices.

Firstly, we stress that even a noiseless quantum computer introduces significant errors
into basic subroutines -- such as the inner product computation -- due to finite sampling.
Nonetheless, as we demonstrated in this simple example, a good optimizer may still find a good solution for
a moderate number of shots.

Secondly, estimating expectation values of differential operators involves a trade-off between the number of circuits
to run and the number of 2-qubit entangling gates in each circuit. A scalable variant of the algorithm necessarily involves a
large number of 2-qubit gates.

Thirdly, the estimation of gradients is costly, at the moment, and does not seem to pay off due to insufficient
precision. This point also crucially depends on the accuracy of differential operator estimations.

Finally, although current error mitigation methods may improve the accuracy of cost function estimation,
it is insufficient to achieve overall good convergence.
We note that there is a certain degree of freedom concerning the choice of tests to present: one could present much more accurate
results for 2 or 3 qubits.
Nonetheless, within the context of NISQ-viability and considering the modern standards
in classical numerical simulation,
we believe the presented results are sufficient to cast serious doubt
about the near-term applicability of quantum computers to PDEs.


\section*{Acknowledgements}
This work was supported by the project AnQuC-3 of the Competence Center Quantum Computing Rhineland-
Palatinate (Germany).
We acknowledge the use of IBM Quantum services for this work. The views expressed are those of the authors, and do not reflect the official policy or position of IBM or the IBM Quantum team.

\section*{Data Availability}
The main tests were performed on
\texttt{ibmq\_ehningen} (processor type Falcon r5.11, version 3.1.21)
and \texttt{ibmq\_montreal} (processor type Falcon r4, version 1.11.26).
The coupling maps are displayed in Fig.~\ref{fig:ibm_layouts}.
The calibration data is detailed in Table~\ref{table:ehningen} and
\ref{table:montreal}.
The \texttt{Qiskit} code can be found in \url{https://github.com/MazenAli/VQA_Poisson1D}.


\bibliography{citations}


\appendix
\newpage

\begin{figure*}[t]
	\centering
	\begin{subfigure}{1.5\columnwidth}
	\centering
	\scalebox{0.5}{
\Qcircuit @C=1.0em @R=0.2em @!R { \\
	 	\nghost{{q}_{0} :  } & \lstick{\ket{0}} & \gate{\mathrm{R_Y}\,(\mathrm{{\ensuremath{\theta}}[0]})} & \ctrl{1} & \gate{\mathrm{R_Y}\,(\mathrm{{\ensuremath{\theta}}[5]})} & \qw & \ctrl{1} & \gate{\mathrm{R_Y}\,(\mathrm{{\ensuremath{\theta}}[10]})} & \qw & \qw & \qw & \qw & \qw\\
	 	\nghost{{q}_{1} :  } & \lstick{\ket{0}} & \gate{\mathrm{R_Y}\,(\mathrm{{\ensuremath{\theta}}[1]})} & \control\qw & \ctrl{1} & \gate{\mathrm{R_Y}\,(\mathrm{{\ensuremath{\theta}}[6]})} & \control\qw & \ctrl{1} & \gate{\mathrm{R_Y}\,(\mathrm{{\ensuremath{\theta}}[11]})} & \qw & \qw & \qw & \qw\\
	 	\nghost{{q}_{2} :  } & \lstick{\ket{0}} & \gate{\mathrm{R_Y}\,(\mathrm{{\ensuremath{\theta}}[2]})} & \qw & \control\qw & \ctrl{1} & \gate{\mathrm{R_Y}\,(\mathrm{{\ensuremath{\theta}}[7]})} & \control\qw & \ctrl{1} & \gate{\mathrm{R_Y}\,(\mathrm{{\ensuremath{\theta}}[12]})} & \qw & \qw & \qw\\
	 	\nghost{{q}_{3} :  } & \lstick{\ket{0}} & \gate{\mathrm{R_Y}\,(\mathrm{{\ensuremath{\theta}}[3]})} & \qw & \qw & \control\qw & \ctrl{1} & \gate{\mathrm{R_Y}\,(\mathrm{{\ensuremath{\theta}}[8]})} & \control\qw & \ctrl{1} & \gate{\mathrm{R_Y}\,(\mathrm{{\ensuremath{\theta}}[13]})} & \qw & \qw\\
	 	\nghost{{q}_{4} :  } & \lstick{\ket{0}} & \gate{\mathrm{R_Y}\,(\mathrm{{\ensuremath{\theta}}[4]})} & \qw & \qw & \qw & \control\qw & \gate{\mathrm{R_Y}\,(\mathrm{{\ensuremath{\theta}}[9]})} & \qw & \control\qw & \gate{\mathrm{R_Y}\,(\mathrm{{\ensuremath{\theta}}[14]})} & \qw & \qw\\
\\ }}
	\caption{Linear $R_Y$-$CZ$, 5 qubits, 2 layers.}
	
	\end{subfigure}\hfill
	\begin{subfigure}{1.5\columnwidth}
	\centering
	\scalebox{0.8}{
\Qcircuit @C=1.0em @R=0.2em @!R { \\
	 	\nghost{{q}_{0} :  } & \lstick{\ket{0}} & \gate{\mathrm{R_Y}\,(\mathrm{{\ensuremath{\theta}}[0]})} & \ctrl{1} & \gate{\mathrm{R_Y}\,(\mathrm{{\ensuremath{\theta}}[5]})} & \qw & \qw & \ctrl{1} & \gate{\mathrm{R_Y}\,(\mathrm{{\ensuremath{\theta}}[13]})} & \qw & \qw & \qw & \qw\\
	 	\nghost{{q}_{1} :  } & \lstick{\ket{0}} & \gate{\mathrm{R_Y}\,(\mathrm{{\ensuremath{\theta}}[1]})} & \control\qw & \gate{\mathrm{R_Y}\,(\mathrm{{\ensuremath{\theta}}[6]})} & \ctrl{1} & \gate{\mathrm{R_Y}\,(\mathrm{{\ensuremath{\theta}}[9]})} & \control\qw & \gate{\mathrm{R_Y}\,(\mathrm{{\ensuremath{\theta}}[14]})} & \ctrl{1} & \gate{\mathrm{R_Y}\,(\mathrm{{\ensuremath{\theta}}[17]})} & \qw & \qw\\
	 	\nghost{{q}_{2} :  } & \lstick{\ket{0}} & \gate{\mathrm{R_Y}\,(\mathrm{{\ensuremath{\theta}}[2]})} & \ctrl{1} & \gate{\mathrm{R_Y}\,(\mathrm{{\ensuremath{\theta}}[7]})} & \control\qw & \gate{\mathrm{R_Y}\,(\mathrm{{\ensuremath{\theta}}[10]})} & \ctrl{1} & \gate{\mathrm{R_Y}\,(\mathrm{{\ensuremath{\theta}}[15]})} & \control\qw & \gate{\mathrm{R_Y}\,(\mathrm{{\ensuremath{\theta}}[18]})} & \qw & \qw\\
	 	\nghost{{q}_{3} :  } & \lstick{\ket{0}} & \gate{\mathrm{R_Y}\,(\mathrm{{\ensuremath{\theta}}[3]})} & \control\qw & \gate{\mathrm{R_Y}\,(\mathrm{{\ensuremath{\theta}}[8]})} & \ctrl{1} & \gate{\mathrm{R_Y}\,(\mathrm{{\ensuremath{\theta}}[11]})} & \control\qw & \gate{\mathrm{R_Y}\,(\mathrm{{\ensuremath{\theta}}[16]})} & \ctrl{1} & \gate{\mathrm{R_Y}\,(\mathrm{{\ensuremath{\theta}}[19]})} & \qw & \qw\\
	 	\nghost{{q}_{4} :  } & \lstick{\ket{0}} & \gate{\mathrm{R_Y}\,(\mathrm{{\ensuremath{\theta}}[4]})} & \qw & \qw & \control\qw & \gate{\mathrm{R_Y}\,(\mathrm{{\ensuremath{\theta}}[12]})} & \qw & \qw & \control\qw & \gate{\mathrm{R_Y}\,(\mathrm{{\ensuremath{\theta}}[20]})} & \qw & \qw\\
\\ }}
	\caption{Linear alternating $R_Y$-$CZ$, 5 qubits, 2 layers.}
	\end{subfigure}\hfill
	\begin{subfigure}{1.5\columnwidth}
	\centering
	\scalebox{0.3}{
\Qcircuit @C=1.0em @R=0.2em @!R { \\
	 	\nghost{{q}_{0} :  } & \lstick{\ket{0}} & \gate{\mathrm{R_X}\,(\mathrm{{\ensuremath{\theta}}[0]})} & \gate{\mathrm{R_Y}\,(\mathrm{{\ensuremath{\theta}}[1]})} & \gate{\mathrm{R_Z}\,(\mathrm{{\ensuremath{\theta}}[2]})} & \ctrl{1} & \gate{\mathrm{R_X}\,(\mathrm{{\ensuremath{\theta}}[15]})} & \gate{\mathrm{R_Y}\,(\mathrm{{\ensuremath{\theta}}[16]})} & \gate{\mathrm{R_Z}\,(\mathrm{{\ensuremath{\theta}}[17]})} & \qw & \qw & \qw & \qw & \ctrl{4} & \gate{\mathrm{R_X}\,(\mathrm{{\ensuremath{\theta}}[39]})} & \gate{\mathrm{R_Y}\,(\mathrm{{\ensuremath{\theta}}[40]})} & \gate{\mathrm{R_Z}\,(\mathrm{{\ensuremath{\theta}}[41]})} & \ctrl{1} & \gate{\mathrm{R_X}\,(\mathrm{{\ensuremath{\theta}}[45]})} & \gate{\mathrm{R_Y}\,(\mathrm{{\ensuremath{\theta}}[46]})} & \gate{\mathrm{R_Z}\,(\mathrm{{\ensuremath{\theta}}[47]})} & \qw & \ctrl{4} & \gate{\mathrm{R_X}\,(\mathrm{{\ensuremath{\theta}}[69]})} & \gate{\mathrm{R_Y}\,(\mathrm{{\ensuremath{\theta}}[70]})} & \gate{\mathrm{R_Z}\,(\mathrm{{\ensuremath{\theta}}[71]})} & \qw & \qw\\
	 	\nghost{{q}_{1} :  } & \lstick{\ket{0}} & \gate{\mathrm{R_X}\,(\mathrm{{\ensuremath{\theta}}[3]})} & \gate{\mathrm{R_Y}\,(\mathrm{{\ensuremath{\theta}}[4]})} & \gate{\mathrm{R_Z}\,(\mathrm{{\ensuremath{\theta}}[5]})} & \targ & \gate{\mathrm{R_X}\,(\mathrm{{\ensuremath{\theta}}[18]})} & \gate{\mathrm{R_Y}\,(\mathrm{{\ensuremath{\theta}}[19]})} & \gate{\mathrm{R_Z}\,(\mathrm{{\ensuremath{\theta}}[20]})} & \ctrl{1} & \gate{\mathrm{R_X}\,(\mathrm{{\ensuremath{\theta}}[27]})} & \gate{\mathrm{R_Y}\,(\mathrm{{\ensuremath{\theta}}[28]})} & \gate{\mathrm{R_Z}\,(\mathrm{{\ensuremath{\theta}}[29]})} & \qw & \qw & \qw & \qw & \targ & \gate{\mathrm{R_X}\,(\mathrm{{\ensuremath{\theta}}[48]})} & \gate{\mathrm{R_Y}\,(\mathrm{{\ensuremath{\theta}}[49]})} & \gate{\mathrm{R_Z}\,(\mathrm{{\ensuremath{\theta}}[50]})} & \ctrl{1} & \qw & \gate{\mathrm{R_X}\,(\mathrm{{\ensuremath{\theta}}[57]})} & \gate{\mathrm{R_Y}\,(\mathrm{{\ensuremath{\theta}}[58]})} & \gate{\mathrm{R_Z}\,(\mathrm{{\ensuremath{\theta}}[59]})} & \qw & \qw\\
	 	\nghost{{q}_{2} :  } & \lstick{\ket{0}} & \gate{\mathrm{R_X}\,(\mathrm{{\ensuremath{\theta}}[6]})} & \gate{\mathrm{R_Y}\,(\mathrm{{\ensuremath{\theta}}[7]})} & \gate{\mathrm{R_Z}\,(\mathrm{{\ensuremath{\theta}}[8]})} & \ctrl{1} & \gate{\mathrm{R_X}\,(\mathrm{{\ensuremath{\theta}}[21]})} & \gate{\mathrm{R_Y}\,(\mathrm{{\ensuremath{\theta}}[22]})} & \gate{\mathrm{R_Z}\,(\mathrm{{\ensuremath{\theta}}[23]})} & \targ & \gate{\mathrm{R_X}\,(\mathrm{{\ensuremath{\theta}}[30]})} & \gate{\mathrm{R_Y}\,(\mathrm{{\ensuremath{\theta}}[31]})} & \gate{\mathrm{R_Z}\,(\mathrm{{\ensuremath{\theta}}[32]})} & \qw & \ctrl{1} & \gate{\mathrm{R_X}\,(\mathrm{{\ensuremath{\theta}}[51]})} & \gate{\mathrm{R_Y}\,(\mathrm{{\ensuremath{\theta}}[52]})} & \gate{\mathrm{R_Z}\,(\mathrm{{\ensuremath{\theta}}[53]})} & \qw & \qw & \qw & \targ & \qw & \gate{\mathrm{R_X}\,(\mathrm{{\ensuremath{\theta}}[60]})} & \gate{\mathrm{R_Y}\,(\mathrm{{\ensuremath{\theta}}[61]})} & \gate{\mathrm{R_Z}\,(\mathrm{{\ensuremath{\theta}}[62]})} & \qw & \qw\\
	 	\nghost{{q}_{3} :  } & \lstick{\ket{0}} & \gate{\mathrm{R_X}\,(\mathrm{{\ensuremath{\theta}}[9]})} & \gate{\mathrm{R_Y}\,(\mathrm{{\ensuremath{\theta}}[10]})} & \gate{\mathrm{R_Z}\,(\mathrm{{\ensuremath{\theta}}[11]})} & \targ & \gate{\mathrm{R_X}\,(\mathrm{{\ensuremath{\theta}}[24]})} & \gate{\mathrm{R_Y}\,(\mathrm{{\ensuremath{\theta}}[25]})} & \gate{\mathrm{R_Z}\,(\mathrm{{\ensuremath{\theta}}[26]})} & \ctrl{1} & \gate{\mathrm{R_X}\,(\mathrm{{\ensuremath{\theta}}[33]})} & \gate{\mathrm{R_Y}\,(\mathrm{{\ensuremath{\theta}}[34]})} & \gate{\mathrm{R_Z}\,(\mathrm{{\ensuremath{\theta}}[35]})} & \qw & \targ & \gate{\mathrm{R_X}\,(\mathrm{{\ensuremath{\theta}}[54]})} & \gate{\mathrm{R_Y}\,(\mathrm{{\ensuremath{\theta}}[55]})} & \gate{\mathrm{R_Z}\,(\mathrm{{\ensuremath{\theta}}[56]})} & \ctrl{1} & \gate{\mathrm{R_X}\,(\mathrm{{\ensuremath{\theta}}[63]})} & \gate{\mathrm{R_Y}\,(\mathrm{{\ensuremath{\theta}}[64]})} & \gate{\mathrm{R_Z}\,(\mathrm{{\ensuremath{\theta}}[65]})} & \qw & \qw & \qw & \qw & \qw & \qw\\
	 	\nghost{{q}_{4} :  } & \lstick{\ket{0}} & \gate{\mathrm{R_X}\,(\mathrm{{\ensuremath{\theta}}[12]})} & \gate{\mathrm{R_Y}\,(\mathrm{{\ensuremath{\theta}}[13]})} & \gate{\mathrm{R_Z}\,(\mathrm{{\ensuremath{\theta}}[14]})} & \qw & \qw & \qw & \qw & \targ & \gate{\mathrm{R_X}\,(\mathrm{{\ensuremath{\theta}}[36]})} & \gate{\mathrm{R_Y}\,(\mathrm{{\ensuremath{\theta}}[37]})} & \gate{\mathrm{R_Z}\,(\mathrm{{\ensuremath{\theta}}[38]})} & \targ & \gate{\mathrm{R_X}\,(\mathrm{{\ensuremath{\theta}}[42]})} & \gate{\mathrm{R_Y}\,(\mathrm{{\ensuremath{\theta}}[43]})} & \gate{\mathrm{R_Z}\,(\mathrm{{\ensuremath{\theta}}[44]})} & \qw & \targ & \gate{\mathrm{R_X}\,(\mathrm{{\ensuremath{\theta}}[66]})} & \gate{\mathrm{R_Y}\,(\mathrm{{\ensuremath{\theta}}[67]})} & \gate{\mathrm{R_Z}\,(\mathrm{{\ensuremath{\theta}}[68]})} & \targ & \gate{\mathrm{R_X}\,(\mathrm{{\ensuremath{\theta}}[72]})} & \gate{\mathrm{R_Y}\,(\mathrm{{\ensuremath{\theta}}[73]})} & \gate{\mathrm{R_Z}\,(\mathrm{{\ensuremath{\theta}}[74]})} & \qw & \qw\\
\\ }}
	\caption{Linear alternating periodic $U3$-$CX$, 5 qubits, 2 layers.}
	\end{subfigure}\hfill
	\begin{subfigure}{1.5\columnwidth}
	\centering
	\scalebox{0.4}{
\Qcircuit @C=1.0em @R=0.2em @!R { \\
	 	\nghost{ {q}_{0} :  } & \lstick{\ket{0}} & \gate{\mathrm{R_Z}\,(\mathrm{{\ensuremath{\theta}}[0]})} & \ctrl{1} & \gate{\mathrm{R_Z}\,(\mathrm{{\ensuremath{\theta}}[5]})} & \qw & \qw & \ctrl{4} & \gate{\mathrm{R_Z}\,(\mathrm{{\ensuremath{\theta}}[13]})} & \ctrl{1} & \gate{\mathrm{R_Z}\,(\mathrm{{\ensuremath{\theta}}[15]})} & \qw & \ctrl{4} & \gate{\mathrm{R_Z}\,(\mathrm{{\ensuremath{\theta}}[23]})} & \targ & \gate{\mathrm{R_Z}\,(\mathrm{{\ensuremath{\theta}}[25]})} & \qw & \qw & \targ & \gate{\mathrm{R_Z}\,(\mathrm{{\ensuremath{\theta}}[33]})} & \targ & \gate{\mathrm{R_Z}\,(\mathrm{{\ensuremath{\theta}}[35]})} & \qw & \targ & \gate{\mathrm{R_Z}\,(\mathrm{{\ensuremath{\theta}}[43]})} & \qw & \qw\\ 
	 	\nghost{ {q}_{1} :  } & \lstick{\ket{0}} & \gate{\mathrm{R_Z}\,(\mathrm{{\ensuremath{\theta}}[1]})} & \targ & \gate{\mathrm{R_Z}\,(\mathrm{{\ensuremath{\theta}}[6]})} & \ctrl{1} & \gate{\mathrm{R_Z}\,(\mathrm{{\ensuremath{\theta}}[9]})} & \qw & \qw & \targ & \gate{\mathrm{R_Z}\,(\mathrm{{\ensuremath{\theta}}[16]})} & \ctrl{1} & \qw & \gate{\mathrm{R_Z}\,(\mathrm{{\ensuremath{\theta}}[19]})} & \ctrl{-1} & \gate{\mathrm{R_Z}\,(\mathrm{{\ensuremath{\theta}}[26]})} & \targ & \gate{\mathrm{R_Z}\,(\mathrm{{\ensuremath{\theta}}[29]})} & \qw & \qw & \ctrl{-1} & \gate{\mathrm{R_Z}\,(\mathrm{{\ensuremath{\theta}}[36]})} & \targ & \qw & \gate{\mathrm{R_Z}\,(\mathrm{{\ensuremath{\theta}}[39]})} & \qw & \qw\\ 
	 	\nghost{ {q}_{2} :  } & \lstick{\ket{0}} & \gate{\mathrm{R_Z}\,(\mathrm{{\ensuremath{\theta}}[2]})} & \ctrl{1} & \gate{\mathrm{R_Z}\,(\mathrm{{\ensuremath{\theta}}[7]})} & \targ & \gate{\mathrm{R_Z}\,(\mathrm{{\ensuremath{\theta}}[10]})} & \qw & \ctrl{1} & \gate{\mathrm{R_Z}\,(\mathrm{{\ensuremath{\theta}}[17]})} & \qw & \targ & \qw & \gate{\mathrm{R_Z}\,(\mathrm{{\ensuremath{\theta}}[20]})} & \targ & \gate{\mathrm{R_Z}\,(\mathrm{{\ensuremath{\theta}}[27]})} & \ctrl{-1} & \gate{\mathrm{R_Z}\,(\mathrm{{\ensuremath{\theta}}[30]})} & \qw & \targ & \gate{\mathrm{R_Z}\,(\mathrm{{\ensuremath{\theta}}[37]})} & \qw & \ctrl{-1} & \qw & \gate{\mathrm{R_Z}\,(\mathrm{{\ensuremath{\theta}}[40]})} & \qw & \qw\\ 
	 	\nghost{ {q}_{3} :  } & \lstick{\ket{0}} & \gate{\mathrm{R_Z}\,(\mathrm{{\ensuremath{\theta}}[3]})} & \targ & \gate{\mathrm{R_Z}\,(\mathrm{{\ensuremath{\theta}}[8]})} & \ctrl{1} & \gate{\mathrm{R_Z}\,(\mathrm{{\ensuremath{\theta}}[11]})} & \qw & \targ & \gate{\mathrm{R_Z}\,(\mathrm{{\ensuremath{\theta}}[18]})} & \ctrl{1} & \gate{\mathrm{R_Z}\,(\mathrm{{\ensuremath{\theta}}[21]})} & \qw & \qw & \ctrl{-1} & \gate{\mathrm{R_Z}\,(\mathrm{{\ensuremath{\theta}}[28]})} & \targ & \gate{\mathrm{R_Z}\,(\mathrm{{\ensuremath{\theta}}[31]})} & \qw & \ctrl{-1} & \gate{\mathrm{R_Z}\,(\mathrm{{\ensuremath{\theta}}[38]})} & \targ & \gate{\mathrm{R_Z}\,(\mathrm{{\ensuremath{\theta}}[41]})} & \qw & \qw & \qw & \qw\\ 
	 	\nghost{ {q}_{4} :  } & \lstick{\ket{0}} & \gate{\mathrm{R_Z}\,(\mathrm{{\ensuremath{\theta}}[4]})} & \qw & \qw & \targ & \gate{\mathrm{R_Z}\,(\mathrm{{\ensuremath{\theta}}[12]})} & \targ & \gate{\mathrm{R_Z}\,(\mathrm{{\ensuremath{\theta}}[14]})} & \qw & \targ & \gate{\mathrm{R_Z}\,(\mathrm{{\ensuremath{\theta}}[22]})} & \targ & \gate{\mathrm{R_Z}\,(\mathrm{{\ensuremath{\theta}}[24]})} & \qw & \qw & \ctrl{-1} & \gate{\mathrm{R_Z}\,(\mathrm{{\ensuremath{\theta}}[32]})} & \ctrl{-4} & \gate{\mathrm{R_Z}\,(\mathrm{{\ensuremath{\theta}}[34]})} & \qw & \ctrl{-1} & \gate{\mathrm{R_Z}\,(\mathrm{{\ensuremath{\theta}}[42]})} & \ctrl{-4} & \gate{\mathrm{R_Z}\,(\mathrm{{\ensuremath{\theta}}[44]})} & \qw & \qw\\ 
\\ }}
	\caption{Linear alternating periodic bidirectional $R_Z$-$CX$, 5 qubits, 2 layers.}
	\end{subfigure}\hfill
	\begin{subfigure}{1.5\columnwidth}
	\centering
	\scalebox{0.6}{
\Qcircuit @C=1.0em @R=0.2em @!R { \\
	 	\nghost{ {q}_{0} :  } & \lstick{\ket{0}} & \gate{\mathrm{H}} & \ctrl{1} & \dstick{\hspace{2.0em}\mathrm{ZZ}\,(\mathrm{{\ensuremath{\theta}}[0]})} \qw & \qw & \qw & \gate{\mathrm{R_X}\,(\mathrm{{\ensuremath{\theta}}[1]})} & \qw & \qw & \qw & \qw & \qw & \qw & \qw & \ctrl{1} & \dstick{\hspace{2.0em}\mathrm{ZZ}\,(\mathrm{{\ensuremath{\theta}}[2]})} \qw & \qw & \qw & \gate{\mathrm{R_X}\,(\mathrm{{\ensuremath{\theta}}[3]})} & \qw & \qw & \qw & \qw & \qw & \qw & \qw & \qw & \qw & \qw & \qw & \qw & \qw & \qw\\ 
	 	\nghost{ {q}_{1} :  } & \lstick{\ket{0}} & \gate{\mathrm{H}} & \control \qw & \qw & \qw & \qw & \ctrl{1} & \dstick{\hspace{2.0em}\mathrm{ZZ}\,(\mathrm{{\ensuremath{\theta}}[0]})} \qw & \qw & \qw & \gate{\mathrm{R_X}\,(\mathrm{{\ensuremath{\theta}}[1]})} & \qw & \qw & \qw & \control \qw & \qw & \qw & \qw & \ctrl{1} & \dstick{\hspace{2.0em}\mathrm{ZZ}\,(\mathrm{{\ensuremath{\theta}}[2]})} \qw & \qw & \qw & \gate{\mathrm{R_X}\,(\mathrm{{\ensuremath{\theta}}[3]})} & \qw & \qw & \qw & \qw & \qw & \qw & \qw & \qw & \qw & \qw\\ 
	 	\nghost{ {q}_{2} :  } & \lstick{\ket{0}} & \gate{\mathrm{H}} & \qw & \qw & \qw & \qw & \control \qw & \qw & \qw & \qw & \ctrl{1} & \dstick{\hspace{2.0em}\mathrm{ZZ}\,(\mathrm{{\ensuremath{\theta}}[0]})} \qw & \qw & \qw & \gate{\mathrm{R_X}\,(\mathrm{{\ensuremath{\theta}}[1]})} & \qw & \qw & \qw & \control \qw & \qw & \qw & \qw & \ctrl{1} & \dstick{\hspace{2.0em}\mathrm{ZZ}\,(\mathrm{{\ensuremath{\theta}}[2]})} \qw & \qw & \qw & \gate{\mathrm{R_X}\,(\mathrm{{\ensuremath{\theta}}[3]})} & \qw & \qw & \qw & \qw & \qw & \qw\\ 
	 	\nghost{ {q}_{3} :  } & \lstick{\ket{0}} & \gate{\mathrm{H}} & \qw & \qw & \qw & \qw & \qw & \qw & \qw & \qw & \control \qw & \qw & \qw & \qw & \ctrl{1} & \dstick{\hspace{2.0em}\mathrm{ZZ}\,(\mathrm{{\ensuremath{\theta}}[0]})} \qw & \qw & \qw & \gate{\mathrm{R_X}\,(\mathrm{{\ensuremath{\theta}}[1]})} & \qw & \qw & \qw & \control \qw & \qw & \qw & \qw & \ctrl{1} & \dstick{\hspace{2.0em}\mathrm{ZZ}\,(\mathrm{{\ensuremath{\theta}}[2]})} \qw & \qw & \qw & \gate{\mathrm{R_X}\,(\mathrm{{\ensuremath{\theta}}[3]})} & \qw & \qw\\ 
	 	\nghost{ {q}_{4} :  } & \lstick{\ket{0}} & \gate{\mathrm{H}} & \qw & \qw & \qw & \qw & \qw & \qw & \qw & \qw & \qw & \qw & \qw & \qw & \control \qw & \qw & \qw & \qw & \gate{\mathrm{R_X}\,(\mathrm{{\ensuremath{\theta}}[1]})} & \qw & \qw & \qw & \qw & \qw & \qw & \qw & \control \qw & \qw & \qw & \qw & \gate{\mathrm{R_X}\,(\mathrm{{\ensuremath{\theta}}[3]})} & \qw & \qw\\ 
\\ }}
	\caption{QAOA (inspired) ansatz, 5 qubits, 2 layers.}
	\end{subfigure}\hfill
	\begin{subfigure}{1.5\columnwidth}
	\centering
	\scalebox{0.35}{
\Qcircuit @C=1.0em @R=0.2em @!R { \\
	 	\nghost{{q}_{0} :  } & \lstick{\ket{0}} & \gate{\mathrm{H}} & \ctrl{1} & \dstick{\hspace{2.0em}\mathrm{ZZ}\,(\mathrm{{\ensuremath{\theta}}[0]})} \qw & \qw & \qw & \gate{\mathrm{R_X}\,(\mathrm{\frac{\pi}{2}})} & \qw & \qw & \qw & \qw & \qw & \qw & \qw & \qw & \qw & \qw & \qw & \qw & \ctrl{4} & \qw & \ctrl{4} & \gate{\mathrm{R_X}\,(\mathrm{\frac{-\pi}{2}})} & \gate{\mathrm{R_X}\,(\mathrm{{\ensuremath{\theta}}[1]})} & \ctrl{1} & \dstick{\hspace{2.0em}\mathrm{ZZ}\,(\mathrm{{\ensuremath{\theta}}[2]})} \qw & \qw & \qw & \gate{\mathrm{R_X}\,(\mathrm{\frac{\pi}{2}})} & \qw & \qw & \qw & \qw & \qw & \qw & \qw & \qw & \qw & \qw & \qw & \qw & \ctrl{4} & \qw & \ctrl{4} & \gate{\mathrm{R_X}\,(\mathrm{\frac{-\pi}{2}})} & \gate{\mathrm{R_X}\,(\mathrm{{\ensuremath{\theta}}[3]})} & \qw & \qw\\
	 	\nghost{{q}_{1} :  } & \lstick{\ket{0}} & \gate{\mathrm{H}} & \control \qw & \qw & \qw & \qw & \ctrl{1} & \dstick{\hspace{2.0em}\mathrm{ZZ}\,(\mathrm{{\ensuremath{\theta}}[0]})} \qw & \qw & \qw & \gate{\mathrm{R_X}\,(\mathrm{{\ensuremath{\theta}}[1]})} & \qw & \qw & \qw & \qw & \qw & \qw & \qw & \qw & \qw & \qw & \qw & \qw & \qw & \control \qw & \qw & \qw & \qw & \ctrl{1} & \dstick{\hspace{2.0em}\mathrm{ZZ}\,(\mathrm{{\ensuremath{\theta}}[2]})} \qw & \qw & \qw & \gate{\mathrm{R_X}\,(\mathrm{{\ensuremath{\theta}}[3]})} & \qw & \qw & \qw & \qw & \qw & \qw & \qw & \qw & \qw & \qw & \qw & \qw & \qw & \qw & \qw\\
	 	\nghost{{q}_{2} :  } & \lstick{\ket{0}} & \gate{\mathrm{H}} & \qw & \qw & \qw & \qw & \control \qw & \qw & \qw & \qw & \ctrl{1} & \dstick{\hspace{2.0em}\mathrm{ZZ}\,(\mathrm{{\ensuremath{\theta}}[0]})} \qw & \qw & \qw & \gate{\mathrm{R_X}\,(\mathrm{{\ensuremath{\theta}}[1]})} & \qw & \qw & \qw & \qw & \qw & \qw & \qw & \qw & \qw & \qw & \qw & \qw & \qw & \control \qw & \qw & \qw & \qw & \ctrl{1} & \dstick{\hspace{2.0em}\mathrm{ZZ}\,(\mathrm{{\ensuremath{\theta}}[2]})} \qw & \qw & \qw & \gate{\mathrm{R_X}\,(\mathrm{{\ensuremath{\theta}}[3]})} & \qw & \qw & \qw & \qw & \qw & \qw & \qw & \qw & \qw & \qw & \qw\\
	 	\nghost{{q}_{3} :  } & \lstick{\ket{0}} & \gate{\mathrm{H}} & \qw & \qw & \qw & \qw & \qw & \qw & \qw & \qw & \control \qw & \qw & \qw & \qw & \ctrl{1} & \dstick{\hspace{2.0em}\mathrm{ZZ}\,(\mathrm{{\ensuremath{\theta}}[0]})} \qw & \qw & \qw & \gate{\mathrm{R_X}\,(\mathrm{{\ensuremath{\theta}}[1]})} & \qw & \qw & \qw & \qw & \qw & \qw & \qw & \qw & \qw & \qw & \qw & \qw & \qw & \control \qw & \qw & \qw & \qw & \ctrl{1} & \dstick{\hspace{2.0em}\mathrm{ZZ}\,(\mathrm{{\ensuremath{\theta}}[2]})} \qw & \qw & \qw & \gate{\mathrm{R_X}\,(\mathrm{{\ensuremath{\theta}}[3]})} & \qw & \qw & \qw & \qw & \qw & \qw & \qw\\
	 	\nghost{{q}_{4} :  } & \lstick{\ket{0}} & \gate{\mathrm{H}} & \qw & \qw & \qw & \qw & \qw & \qw & \qw & \qw & \qw & \qw & \qw & \qw & \control \qw & \qw & \qw & \qw & \gate{\mathrm{R_X}\,(\mathrm{\frac{\pi}{2}})} & \targ & \gate{\mathrm{R_Z}\,(\mathrm{{\ensuremath{\theta}}[0]})} & \targ & \gate{\mathrm{R_X}\,(\mathrm{\frac{-\pi}{2}})} & \gate{\mathrm{R_X}\,(\mathrm{{\ensuremath{\theta}}[1]})} & \qw & \qw & \qw & \qw & \qw & \qw & \qw & \qw & \qw & \qw & \qw & \qw & \control \qw & \qw & \qw & \qw & \gate{\mathrm{R_X}\,(\mathrm{\frac{\pi}{2}})} & \targ & \gate{\mathrm{R_Z}\,(\mathrm{{\ensuremath{\theta}}[2]})} & \targ & \gate{\mathrm{R_X}\,(\mathrm{\frac{-\pi}{2}})} & \gate{\mathrm{R_X}\,(\mathrm{{\ensuremath{\theta}}[3]})} & \qw & \qw\\
\\ }}
	\caption{QAOA periodic, 5 qubits, 2 layers.}
	\end{subfigure}
	\caption{Ansaetze used for $\ket{\psi(\theta)}$ in this work.}
	\label{fig:ansaetze}
\end{figure*}

\begin{figure*}[t]
	\centering
	\begin{subfigure}{1.5\columnwidth}
	\centering
	\scalebox{1}{
\Qcircuit @C=1.0em @R=0.2em @!R { \\
	 	\nghost{{q}_{0} :  } & \lstick{\ket{0}} & \qw & \multigate{4}{U(\theta)} & \qw & \multigate{4}{U_f} & \qw & \qw & \qw & \qw\\
	 	\nghost{{q}_{1} :  } & \lstick{\ket{0}} & \qw & \ghost{\ket{\psi(\theta)}} & \qw & \ghost{\ket{\bs f}} & \qw & \qw & \qw & \qw\\
	 	\nghost{{q}_{2} :  } & \lstick{\ket{0}} & \qw & \ghost{\ket{\psi(\theta)}} & \qw & \ghost{\ket{\bs f}} & \qw & \qw & \qw & \qw\\
	 	\nghost{{q}_{3} :  } & \lstick{\ket{0}} & \qw & \ghost{\ket{\psi(\theta)}} & \qw & \ghost{\ket{\bs f}} & \qw & \qw & \qw & \qw\\
	 	\nghost{{q}_{4} :  } & \lstick{\ket{0}} & \qw & \ghost{\ket{\psi(\theta)}} & \qw & \ghost{\ket{\bs f}} & \qw & \qw & \qw & \qw\\
	 	\nghost{{q}_{5} :  } & \lstick{\ket{0}} & \gate{\mathrm{H}} & \ctrl{-1} & \gate{\mathrm{X}} & \ctrl{-1} & \gate{\mathrm{H}} & \meter & \qw & \qw\\
	 	\nghost{\mathrm{{c} :  }} & \lstick{\mathrm{{c} :  }} & \lstick{/_{_{1}}} \cw & \cw & \cw & \cw & \cw & \dstick{_{_{\hspace{0.0em}0}}} \cw \ar @{<=} [-1,0] & \cw & \cw\\
\\ }}
	\caption{Hadamard test for estimating $\Re\braket{\bs f|\psi(\theta)}$.}
	
	\end{subfigure}\hfill
	
	\begin{subfigure}{1.5\columnwidth}
	\centering
	\scalebox{1}{
\Qcircuit @C=1.0em @R=0.2em @!R { \\
	 	\nghost{{q}_{0} :  } & \lstick{\ket{0}} & \multigate{4}{U(\theta)} & \multigate{4}{U_f^\dag} & \qw \barrier[0em]{4} & \qw & \meter & \qw & \qw & \qw & \qw & \qw & \qw\\
	 	\nghost{{q}_{1} :  } & \lstick{\ket{0}} & \ghost{\mathrm{R_Y}\,(\mathrm{{\ensuremath{\theta}}[1]})} & \ghost{\mathrm{H}} & \qw & \qw & \qw & \meter & \qw & \qw & \qw & \qw & \qw\\
	 	\nghost{{q}_{2} :  } & \lstick{\ket{0}} & \ghost{\mathrm{R_Y}\,(\mathrm{{\ensuremath{\theta}}[2]})} & \ghost{\mathrm{H}} & \qw & \qw & \qw & \qw & \meter & \qw & \qw & \qw & \qw\\
	 	\nghost{{q}_{3} :  } & \lstick{\ket{0}} & \ghost{\mathrm{R_Y}\,(\mathrm{{\ensuremath{\theta}}[3]})} & \ghost{\mathrm{H}} & \qw & \qw & \qw & \qw & \qw & \meter & \qw & \qw & \qw\\
	 	\nghost{{q}_{4} :  } & \lstick{\ket{0}} & \ghost{\mathrm{R_Y}\,(\mathrm{{\ensuremath{\theta}}[4]})} & \ghost{\mathrm{H}} & \qw & \qw & \qw & \qw & \qw & \qw & \meter & \qw & \qw\\
	 	\nghost{\mathrm{{meas} :  }} & \lstick{\mathrm{{meas} :  }} & \lstick{/_{_{5}}} \cw & \cw & \cw & \cw & \dstick{_{_{\hspace{0.0em}0}}} \cw \ar @{<=} [-5,0] & \dstick{_{_{\hspace{0.0em}1}}} \cw \ar @{<=} [-4,0] & \dstick{_{_{\hspace{0.0em}2}}} \cw \ar @{<=} [-3,0] & \dstick{_{_{\hspace{0.0em}3}}} \cw \ar @{<=} [-2,0] & \dstick{_{_{\hspace{0.0em}4}}} \cw \ar @{<=} [-1,0] & \cw & \cw\\
\\ }}
	\caption{Overlap test for estimating $|\braket{\bs f|\psi(\theta)}|^2$.}
	
	\end{subfigure}
	\caption{Circuits for estimating inner products with
	$\ket{\psi(\theta)}=U(\theta)\ket{0}$ and $\ket{\bs f}=U_{f}\ket{0}$.}
	\label{fig:innerp}
\end{figure*}

\begin{figure*}[t]
	\centering
	\scalebox{1}{
\Qcircuit @C=1.0em @R=0.2em @!R { \\
	 	\nghost{{q}_{0} :  } & \lstick{{q}_{0} :  } & \gate{\mathrm{H}} & \meter & \qw & \qw\\
	 	\nghost{{q}_{1} :  } & \lstick{{q}_{1} :  } & \qw & \qw & \qw & \qw\\
	 	\nghost{{q}_{2} :  } & \lstick{{q}_{2} :  } & \qw & \qw & \qw & \qw\\
	 	\nghost{{q}_{3} :  } & \lstick{{q}_{3} :  } & \qw & \qw & \qw & \qw\\
	 	\nghost{{q}_{4} :  } & \lstick{{q}_{4} :  } & \qw & \qw & \qw & \qw\\
	 	\nghost{\mathrm{{c} :  }} & \lstick{\mathrm{{c} :  }} & \lstick{/_{_{1}}} \cw & \dstick{_{_{\hspace{0.0em}0}}} \cw \ar @{<=} [-5,0] & \cw & \cw\\
\\ }}
	\caption{Circuit for $I^{\otimes n-1}\otimes X$ required for all decompositions.}
	\label{fig:IX}
\end{figure*}
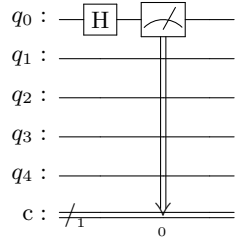

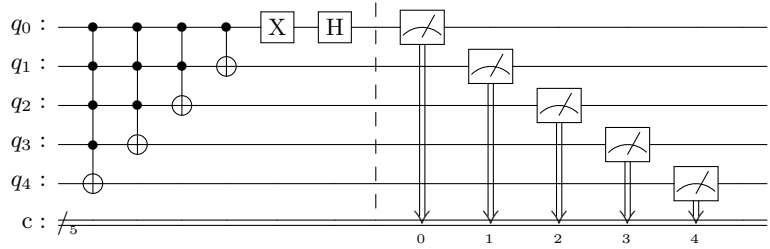
\begin{figure*}[t]
	\centering
	\scalebox{1}{
\Qcircuit @C=1.0em @R=0.2em @!R { \\
	 	\nghost{{q}_{0} :  } & \lstick{{q}_{0} :  } & \ctrl{1} & \ctrl{1} & \ctrl{1} & \ctrl{1} & \gate{\mathrm{X}} & \gate{\mathrm{H}} \barrier[0em]{4} & \qw & \meter & \qw & \qw & \qw & \qw & \qw & \qw\\
	 	\nghost{{q}_{1} :  } & \lstick{{q}_{1} :  } & \ctrl{1} & \ctrl{1} & \ctrl{1} & \targ & \qw & \qw & \qw & \qw & \meter & \qw & \qw & \qw & \qw & \qw\\
	 	\nghost{{q}_{2} :  } & \lstick{{q}_{2} :  } & \ctrl{1} & \ctrl{1} & \targ & \qw & \qw & \qw & \qw & \qw & \qw & \meter & \qw & \qw & \qw & \qw\\
	 	\nghost{{q}_{3} :  } & \lstick{{q}_{3} :  } & \ctrl{1} & \targ & \qw & \qw & \qw & \qw & \qw & \qw & \qw & \qw & \meter & \qw & \qw & \qw\\
	 	\nghost{{q}_{4} :  } & \lstick{{q}_{4} :  } & \targ & \qw & \qw & \qw & \qw & \qw & \qw & \qw & \qw & \qw & \qw & \meter & \qw & \qw\\
	 	\nghost{\mathrm{{c} :  }} & \lstick{\mathrm{{c} :  }} & \lstick{/_{_{5}}} \cw & \cw & \cw & \cw & \cw & \cw & \cw & \dstick{_{_{\hspace{0.0em}0}}} \cw \ar @{<=} [-5,0] & \dstick{_{_{\hspace{0.0em}1}}} \cw \ar @{<=} [-4,0] & \dstick{_{_{\hspace{0.0em}2}}} \cw \ar @{<=} [-3,0] & \dstick{_{_{\hspace{0.0em}3}}} \cw \ar @{<=} [-2,0] & \dstick{_{_{\hspace{0.0em}4}}} \cw \ar @{<=} [-1,0] & \cw & \cw\\
\\ }}
	\caption{Sato21: circuit for estimating expectation of $\bs A$ as in \cite{Sato21}.}
	\label{fig:Sato21}
\end{figure*}

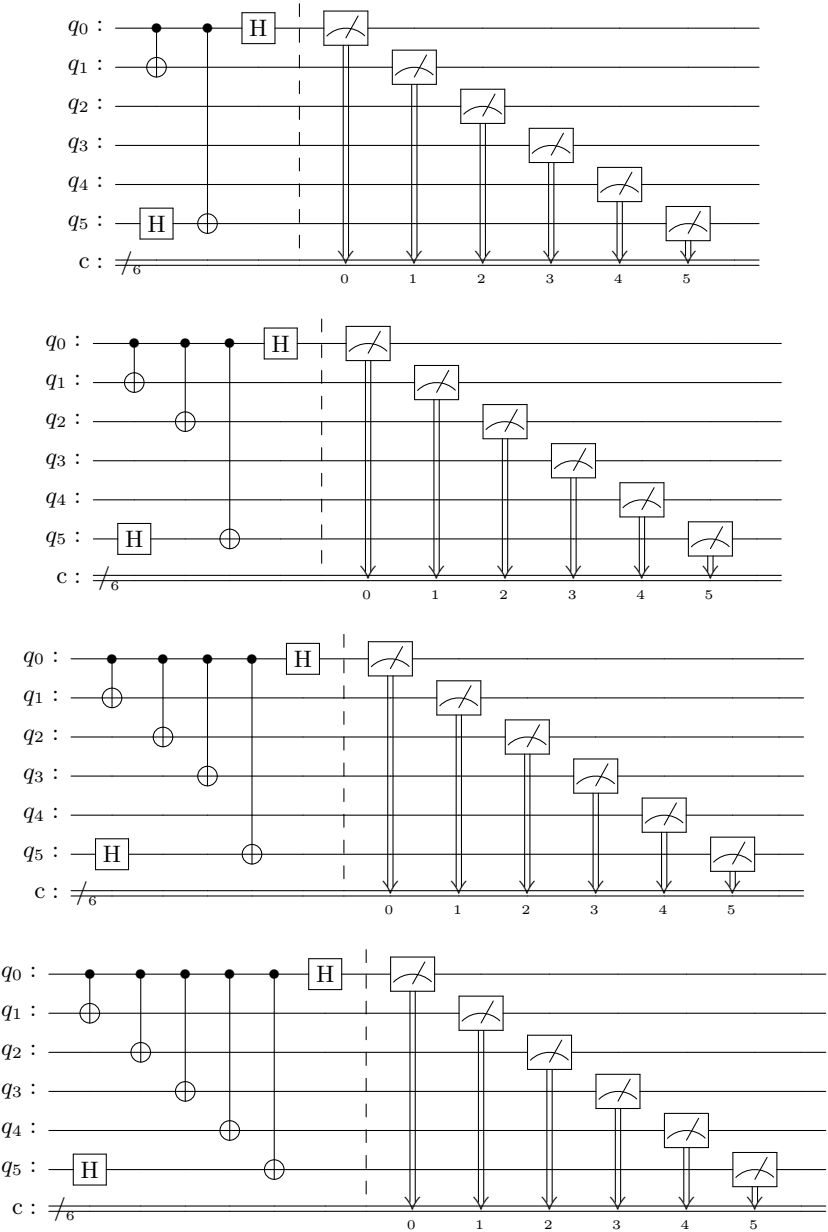
\begin{figure*}[t]
	\centering
	\begin{subfigure}{1.5\columnwidth}
	\centering
	\scalebox{1}{
\Qcircuit @C=1.0em @R=0.2em @!R { \\
	 	\nghost{{q}_{0} :  } & \lstick{{q}_{0} :  } & \ctrl{1} & \ctrl{5} & \gate{\mathrm{H}} \barrier[0em]{5} & \qw & \meter & \qw & \qw & \qw & \qw & \qw & \qw & \qw\\
	 	\nghost{{q}_{1} :  } & \lstick{{q}_{1} :  } & \targ & \qw & \qw & \qw & \qw & \meter & \qw & \qw & \qw & \qw & \qw & \qw\\
	 	\nghost{{q}_{2} :  } & \lstick{{q}_{2} :  } & \qw & \qw & \qw & \qw & \qw & \qw & \meter & \qw & \qw & \qw & \qw & \qw\\
	 	\nghost{{q}_{3} :  } & \lstick{{q}_{3} :  } & \qw & \qw & \qw & \qw & \qw & \qw & \qw & \meter & \qw & \qw & \qw & \qw\\
	 	\nghost{{q}_{4} :  } & \lstick{{q}_{4} :  } & \qw & \qw & \qw & \qw & \qw & \qw & \qw & \qw & \meter & \qw & \qw & \qw\\
	 	\nghost{{q}_{5} :  } & \lstick{{q}_{5} :  } & \gate{\mathrm{H}} & \targ & \qw & \qw & \qw & \qw & \qw & \qw & \qw & \meter & \qw & \qw\\
	 	\nghost{\mathrm{{meas} :  }} & \lstick{\mathrm{{c} :  }} & \lstick{/_{_{6}}} \cw & \cw & \cw & \cw & \dstick{_{_{\hspace{0.0em}0}}} \cw \ar @{<=} [-6,0] & \dstick{_{_{\hspace{0.0em}1}}} \cw \ar @{<=} [-5,0] & \dstick{_{_{\hspace{0.0em}2}}} \cw \ar @{<=} [-4,0] & \dstick{_{_{\hspace{0.0em}3}}} \cw \ar @{<=} [-3,0] & \dstick{_{_{\hspace{0.0em}4}}} \cw \ar @{<=} [-2,0] & \dstick{_{_{\hspace{0.0em}5}}} \cw \ar @{<=} [-1,0] & \cw & \cw\\
\\ }}
	\end{subfigure}\hfill
	\begin{subfigure}{1.5\columnwidth}
	\centering
	\scalebox{1}{
\Qcircuit @C=1.0em @R=0.2em @!R { \\
	 	\nghost{{q}_{0} :  } & \lstick{{q}_{0} :  } & \ctrl{1} & \ctrl{2} & \ctrl{5} & \gate{\mathrm{H}} \barrier[0em]{5} & \qw & \meter & \qw & \qw & \qw & \qw & \qw & \qw & \qw\\
	 	\nghost{{q}_{1} :  } & \lstick{{q}_{1} :  } & \targ & \qw & \qw & \qw & \qw & \qw & \meter & \qw & \qw & \qw & \qw & \qw & \qw\\
	 	\nghost{{q}_{2} :  } & \lstick{{q}_{2} :  } & \qw & \targ & \qw & \qw & \qw & \qw & \qw & \meter & \qw & \qw & \qw & \qw & \qw\\
	 	\nghost{{q}_{3} :  } & \lstick{{q}_{3} :  } & \qw & \qw & \qw & \qw & \qw & \qw & \qw & \qw & \meter & \qw & \qw & \qw & \qw\\
	 	\nghost{{q}_{4} :  } & \lstick{{q}_{4} :  } & \qw & \qw & \qw & \qw & \qw & \qw & \qw & \qw & \qw & \meter & \qw & \qw & \qw\\
	 	\nghost{{q}_{5} :  } & \lstick{{q}_{5} :  } & \gate{\mathrm{H}} & \qw & \targ & \qw & \qw & \qw & \qw & \qw & \qw & \qw & \meter & \qw & \qw\\
	 	\nghost{\mathrm{{meas} :  }} & \lstick{\mathrm{{c} :  }} & \lstick{/_{_{6}}} \cw & \cw & \cw & \cw & \cw & \dstick{_{_{\hspace{0.0em}0}}} \cw \ar @{<=} [-6,0] & \dstick{_{_{\hspace{0.0em}1}}} \cw \ar @{<=} [-5,0] & \dstick{_{_{\hspace{0.0em}2}}} \cw \ar @{<=} [-4,0] & \dstick{_{_{\hspace{0.0em}3}}} \cw \ar @{<=} [-3,0] & \dstick{_{_{\hspace{0.0em}4}}} \cw \ar @{<=} [-2,0] & \dstick{_{_{\hspace{0.0em}5}}} \cw \ar @{<=} [-1,0] & \cw & \cw\\
\\ }}
	\end{subfigure}\hfill
	\begin{subfigure}{1.5\columnwidth}
	\centering
	\scalebox{1}{
\Qcircuit @C=1.0em @R=0.2em @!R { \\
	 	\nghost{{q}_{0} :  } & \lstick{{q}_{0} :  } & \ctrl{1} & \ctrl{2} & \ctrl{3} & \ctrl{5} & \gate{\mathrm{H}} \barrier[0em]{5} & \qw & \meter & \qw & \qw & \qw & \qw & \qw & \qw & \qw\\
	 	\nghost{{q}_{1} :  } & \lstick{{q}_{1} :  } & \targ & \qw & \qw & \qw & \qw & \qw & \qw & \meter & \qw & \qw & \qw & \qw & \qw & \qw\\
	 	\nghost{{q}_{2} :  } & \lstick{{q}_{2} :  } & \qw & \targ & \qw & \qw & \qw & \qw & \qw & \qw & \meter & \qw & \qw & \qw & \qw & \qw\\
	 	\nghost{{q}_{3} :  } & \lstick{{q}_{3} :  } & \qw & \qw & \targ & \qw & \qw & \qw & \qw & \qw & \qw & \meter & \qw & \qw & \qw & \qw\\
	 	\nghost{{q}_{4} :  } & \lstick{{q}_{4} :  } & \qw & \qw & \qw & \qw & \qw & \qw & \qw & \qw & \qw & \qw & \meter & \qw & \qw & \qw\\
	 	\nghost{{q}_{5} :  } & \lstick{{q}_{5} :  } & \gate{\mathrm{H}} & \qw & \qw & \targ & \qw & \qw & \qw & \qw & \qw & \qw & \qw & \meter & \qw & \qw\\
	 	\nghost{\mathrm{{meas} :  }} & \lstick{\mathrm{{c} :  }} & \lstick{/_{_{6}}} \cw & \cw & \cw & \cw & \cw & \cw & \dstick{_{_{\hspace{0.0em}0}}} \cw \ar @{<=} [-6,0] & \dstick{_{_{\hspace{0.0em}1}}} \cw \ar @{<=} [-5,0] & \dstick{_{_{\hspace{0.0em}2}}} \cw \ar @{<=} [-4,0] & \dstick{_{_{\hspace{0.0em}3}}} \cw \ar @{<=} [-3,0] & \dstick{_{_{\hspace{0.0em}4}}} \cw \ar @{<=} [-2,0] & \dstick{_{_{\hspace{0.0em}5}}} \cw \ar @{<=} [-1,0] & \cw & \cw\\
\\ }}
	\end{subfigure}\hfill
	\begin{subfigure}{1.5\columnwidth}
	\centering
	\scalebox{1}{
\Qcircuit @C=1.0em @R=0.2em @!R { \\
	 	\nghost{{q}_{0} :  } & \lstick{{q}_{0} :  } & \ctrl{1} & \ctrl{2} & \ctrl{3} & \ctrl{4} & \ctrl{5} & \gate{\mathrm{H}} \barrier[0em]{5} & \qw & \meter & \qw & \qw & \qw & \qw & \qw & \qw & \qw\\
	 	\nghost{{q}_{1} :  } & \lstick{{q}_{1} :  } & \targ & \qw & \qw & \qw & \qw & \qw & \qw & \qw & \meter & \qw & \qw & \qw & \qw & \qw & \qw\\
	 	\nghost{{q}_{2} :  } & \lstick{{q}_{2} :  } & \qw & \targ & \qw & \qw & \qw & \qw & \qw & \qw & \qw & \meter & \qw & \qw & \qw & \qw & \qw\\
	 	\nghost{{q}_{3} :  } & \lstick{{q}_{3} :  } & \qw & \qw & \targ & \qw & \qw & \qw & \qw & \qw & \qw & \qw & \meter & \qw & \qw & \qw & \qw\\
	 	\nghost{{q}_{4} :  } & \lstick{{q}_{4} :  } & \qw & \qw & \qw & \targ & \qw & \qw & \qw & \qw & \qw & \qw & \qw & \meter & \qw & \qw & \qw\\
	 	\nghost{{q}_{5} :  } & \lstick{{q}_{5} :  } & \gate{\mathrm{H}} & \qw & \qw & \qw & \targ & \qw & \qw & \qw & \qw & \qw & \qw & \qw & \meter & \qw & \qw\\
	 	\nghost{\mathrm{{meas} :  }} & \lstick{\mathrm{{c} :  }} & \lstick{/_{_{6}}} \cw & \cw & \cw & \cw & \cw & \cw & \cw & \dstick{_{_{\hspace{0.0em}0}}} \cw \ar @{<=} [-6,0] & \dstick{_{_{\hspace{0.0em}1}}} \cw \ar @{<=} [-5,0] & \dstick{_{_{\hspace{0.0em}2}}} \cw \ar @{<=} [-4,0] & \dstick{_{_{\hspace{0.0em}3}}} \cw \ar @{<=} [-3,0] & \dstick{_{_{\hspace{0.0em}4}}} \cw \ar @{<=} [-2,0] & \dstick{_{_{\hspace{0.0em}5}}} \cw \ar @{<=} [-1,0] & \cw & \cw\\
\\ }}
	\end{subfigure}
	\caption{Liu21: circuits for estimating expectation of $\bs A$ as in \cite{Liu21}.}
	\label{fig:Liu21}
\end{figure*}

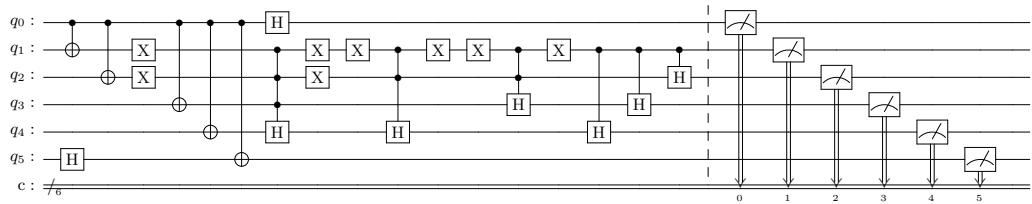
\begin{figure*}[t]
	\centering
	\scalebox{0.7}{
\Qcircuit @C=1.0em @R=0.2em @!R { \\
	 	\nghost{{q}_{0} :  } & \lstick{{q}_{0} :  } & \ctrl{1} & \ctrl{2} & \qw & \ctrl{3} & \ctrl{4} & \ctrl{5} & \gate{\mathrm{H}} & \qw & \qw & \qw & \qw & \qw & \qw & \qw & \qw & \qw & \qw \barrier[0em]{5} & \qw & \meter & \qw & \qw & \qw & \qw & \qw & \qw & \qw\\
	 	\nghost{{q}_{1} :  } & \lstick{{q}_{1} :  } & \targ & \qw & \gate{\mathrm{X}} & \qw & \qw & \qw & \ctrl{1} & \gate{\mathrm{X}} & \gate{\mathrm{X}} & \ctrl{1} & \gate{\mathrm{X}} & \gate{\mathrm{X}} & \ctrl{1} & \gate{\mathrm{X}} & \ctrl{3} & \ctrl{2} & \ctrl{1} & \qw & \qw & \meter & \qw & \qw & \qw & \qw & \qw & \qw\\
	 	\nghost{{q}_{2} :  } & \lstick{{q}_{2} :  } & \qw & \targ & \gate{\mathrm{X}} & \qw & \qw & \qw & \ctrl{1} & \gate{\mathrm{X}} & \qw & \ctrl{2} & \qw & \qw & \ctrl{1} & \qw & \qw & \qw & \gate{\mathrm{H}} & \qw & \qw & \qw & \meter & \qw & \qw & \qw & \qw & \qw\\
	 	\nghost{{q}_{3} :  } & \lstick{{q}_{3} :  } & \qw & \qw & \qw & \targ & \qw & \qw & \ctrl{1} & \qw & \qw & \qw & \qw & \qw & \gate{\mathrm{H}} & \qw & \qw & \gate{\mathrm{H}} & \qw & \qw & \qw & \qw & \qw & \meter & \qw & \qw & \qw & \qw\\
	 	\nghost{{q}_{4} :  } & \lstick{{q}_{4} :  } & \qw & \qw & \qw & \qw & \targ & \qw & \gate{\mathrm{H}} & \qw & \qw & \gate{\mathrm{H}} & \qw & \qw & \qw & \qw & \gate{\mathrm{H}} & \qw & \qw & \qw & \qw & \qw & \qw & \qw & \meter & \qw & \qw & \qw\\
	 	\nghost{{q}_{5} :  } & \lstick{{q}_{5} :  } & \gate{\mathrm{H}} & \qw & \qw & \qw & \qw & \targ & \qw & \qw & \qw & \qw & \qw & \qw & \qw & \qw & \qw & \qw & \qw & \qw & \qw & \qw & \qw & \qw & \qw & \meter & \qw & \qw\\
	 	\nghost{\mathrm{{meas} :  }} & \lstick{\mathrm{{c} :  }} & \lstick{/_{_{6}}} \cw & \cw & \cw & \cw & \cw & \cw & \cw & \cw & \cw & \cw & \cw & \cw & \cw & \cw & \cw & \cw & \cw & \cw & \dstick{_{_{\hspace{0.0em}0}}} \cw \ar @{<=} [-6,0] & \dstick{_{_{\hspace{0.0em}1}}} \cw \ar @{<=} [-5,0] & \dstick{_{_{\hspace{0.0em}2}}} \cw \ar @{<=} [-4,0] & \dstick{_{_{\hspace{0.0em}3}}} \cw \ar @{<=} [-3,0] & \dstick{_{_{\hspace{0.0em}4}}} \cw \ar @{<=} [-2,0] & \dstick{_{_{\hspace{0.0em}5}}} \cw \ar @{<=} [-1,0] & \cw & \cw\\
\\ }}
	\caption{Liu21Grouped: circuit for rotating into the common eigenbasis of the commuting
	operators proposed in \cite{Liu21}.}
	\label{fig:Liu21Grouped}
\end{figure*}

\begin{figure*}[t]
	\centering
	\begin{subfigure}{1.5\columnwidth}
		\centering
		\includegraphics[width=1\columnwidth]{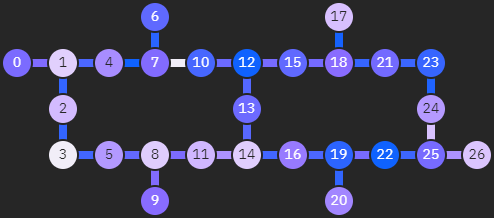}
		\caption{}
	\end{subfigure}\hfill
	\begin{subfigure}{1.5\columnwidth}
		\centering
		\includegraphics[width=1\columnwidth]{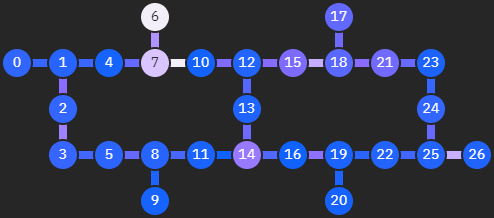}
		\caption{}
	\end{subfigure}
	\caption{IBM backend connectivity maps
	\texttt{ibmq\_ehningen} (a) and \texttt{ibmq\_montreal (b)}.
	Lighter nodes/connections means a larger error.}
	\label{fig:ibm_layouts}
\end{figure*}

\begin{sidewaystable*}
\scalebox{0.35}{
\begin{tabular}{|l|l|l|l|l|l|l|l|l|l|l|l|l|l|}%
    \bfseries Qubit & \bfseries T1(us) & \bfseries T2(us) & \bfseries Frequency (GHz) & \bfseries Anharmonicity (GHz) & \bfseries Readout assignment error & \bfseries Prob meas0 prep1 & \bfseries Prob meas1 prep0 & \bfseries Readout length (ns) & \bfseries ID error & \bfseries sx error & \bfseries Pauli-$X$ error & \bfseries CNOT error & \bfseries Gate time (ns)
	\begin{filecontents*}{ibmq_ehningen_calibrations_2022-12-02T13_40_10Z.csv}
Qubit,T1 (us),T2 (us),Frequency (GHz),Anharmonicity (GHz),Readout assignment error ,Prob meas0 prep1 ,Prob meas1 prep0 ,Readout length (ns),ID error ,asx (sx) error ,Pauli-X error ,CNOT error ,Gate time (ns);;;;
0,50.99240683270166,185.68841503832334,4.960997923991528,-0.3445241971547628,0.007800000000000029,0.0074,0.008199999999999985,846.2222222222222,0.00014898039788878884,0.00014898039788878884,0.00014898039788878884,0\_1:0.005152644673167811,0\_1:352;;;;
1,186.4237175520544,212.8980476253998,5.181915870554795,-0.3400425781524352,0.009500000000000064,0.010800000000000032,0.0082,846.2222222222222,0.0002218683347657844,0.0002218683347657844,0.0002218683347657844,1\_0:0.005152644673167811; 1\_4:0.004712280684157716; 1\_2:0.009166474928205104,1\_0:320; 1\_4:245.33333333333331; 1\_2:263.1111111111111
2,81.8544963052844,9.131407155462284,5.126925983720086,-0.34028204236489296,0.008099999999999996,0.009,0.007199999999999984,846.2222222222222,0.0006089056917463167,0.0006089056917463167,0.0006089056917463167,2\_3:0.009371755500067497; 2\_1:0.009166474928205104,2\_3:352; 2\_1:327.1111111111111;;
3,74.90694045417916,42.23612705801286,5.268148656305129,-0.33842641189959133,0.009200000000000097,0.0098,0.008600000000000052,846.2222222222222,0.00022143637683328093,0.00022143637683328093,0.00022143637683328093,3\_2:0.009371755500067497; 3\_5:0.007968632189730124,3\_2:288; 3\_5:181.33333333333331;;
4,155.24202221100907,167.24230282522404,5.053554814981538,-0.34263703393137723,0.008499999999999952,0.011199999999999988,0.0058,846.2222222222222,0.00022478378392899726,0.00022478378392899726,0.00022478378392899726,4\_1:0.004712280684157716; 4\_7:0.00550868598787127,4\_1:309.3333333333333; 4\_7:202.66666666666666;;
5,91.33674481453554,8.97619774076446,5.071100342978886,-0.34187399904081395,0.009499999999999953,0.0116,0.007399999999999962,846.2222222222222,0.00024219102297458122,0.00024219102297458122,0.00024219102297458122,5\_8:0.009250298229750903; 5\_3:0.007968632189730124,5\_8:273.77777777777777; 5\_3:245.33333333333331;;
6,198.7493313261211,255.154634018372,4.890054468665451,-0.34475347320931227,0.011900000000000022,0.014,0.009800000000000031,846.2222222222222,0.00012256674167892637,0.00012256674167892637,0.00012256674167892637,6\_7:0.008522913672589144,6\_7:309.3333333333333;;;;
7,104.66299988329246,178.8077157995316,4.977682841039246,-0.34397908532965765,0.006699999999999928,0.008,0.00539999999999996,846.2222222222222,0.00013559671330570347,0.00013559671330570347,0.00013559671330570347,7\_10:0.013836010708465463; 7\_6:0.008522913672589144; 7\_4:0.00550868598787127,7\_10:256; 7\_6:245.33333333333331; 7\_4:266.66666666666663
8,93.09228744064546,146.5592338072675,5.174186058948461,-0.33988075897287184,0.017000000000000015,0.019,0.015000000000000013,846.2222222222222,0.0003881828478182781,0.0003881828478182781,0.0003881828478182781,8\_5:0.009250298229750903; 8\_11:0.01090201876377514; 8\_9:0.009488811059412211,8\_5:241.77777777777777; 8\_11:327.1111111111111; 8\_9:391.1111111111111
9,134.39516248980334,144.65069186685818,4.992604184849,-0.34414549718642096,0.012399999999999967,0.01739999999999997,0.0074,846.2222222222222,0.00017317507952390053,0.00017317507952390053,0.00017317507952390053,9\_8:0.009488811059412211,9\_8:423.1111111111111;;;;
10,147.63654069479526,72.67317941699028,4.835220388227256,-0.34705971517657,0.008299999999999974,0.010199999999999987,0.0064,846.2222222222222,0.00019174605488026967,0.00019174605488026967,0.00019174605488026967,10\_7:0.013836010708465463; 10\_12:0.00601257299190297,10\_7:320; 10\_12:259.55555555555554;;
11,127.75982362661964,204.13664287497443,5.119434603125043,-0.34054538758328146,0.017800000000000038,0.0206,0.015000000000000013,846.2222222222222,0.0002940226149296246,0.0002940226149296246,0.0002940226149296246,11\_14:0.008578116707576855; 11\_8:0.01090201876377514,11\_14:352; 11\_8:359.1111111111111;;
12,265.7694631720773,282.3851151034577,4.725482948633703,-0.3484333474508668,0.012299999999999978,0.014399999999999968,0.0102,846.2222222222222,0.00015401832345672973,0.00015401832345672973,0.00015401832345672973,12\_15:0.01030221725514388; 12\_10:0.00601257299190297; 12\_13:0.006548795528996093,12\_15:554.6666666666666; 12\_10:323.55555555555554; 12\_13:252.44444444444443
13,187.69453494449758,170.0872984789343,4.925985534239945,-0.34395637256656786,0.00770000000000004,0.008199999999999985,0.0072,846.2222222222222,0.00015587488620965434,0.00015587488620965434,0.00015587488620965434,13\_12:0.006548795528996093; 13\_14:0.005745084132633443,13\_12:188.44444444444443; 13\_14:263.1111111111111;;
14,183.0709245143768,263.9536804444212,5.176711021127982,-0.3408299348208466,0.006599999999999939,0.009800000000000031,0.0034,846.2222222222222,0.0003404717119728896,0.0003404717119728896,0.0003404717119728896,14\_16:0.006131200810205323; 14\_11:0.008578116707576855; 14\_13:0.005745084132633443,14\_16:440.88888888888886; 14\_11:320; 14\_13:199.1111111111111
15,39.785184316953135,37.676950358006245,4.8928871921304316,-0.3578937985885439,0.008299999999999974,0.0124,0.0041999999999999815,846.2222222222222,0.0017223999797030768,0.0017223999797030768,0.0017223999797030768,15\_12:0.01030221725514388; 15\_18:0.014083120311514485,15\_12:586.6666666666666; 15\_18:444.4444444444444;;
16,157.08225267283365,226.56985120862282,5.022197860887629,-0.34353477301055463,0.005600000000000049,0.006399999999999961,0.0048,846.2222222222222,0.00016595762852781437,0.00016595762852781437,0.00016595762852781437,16\_14:0.006131200810205323; 16\_19:0.006167191319418619,16\_14:472.88888888888886; 16\_19:220.44444444444443;;
17,6.972985306357658,16.105905976155135,5.135638173849081,-0.3410107828456787,0.027900000000000036,0.04459999999999997,0.0112,846.2222222222222,0.0007795879481897243,0.0007795879481897243,0.0007795879481897243,17\_18:0.009878420587822734,17\_18:341.3333333333333;;;;
18,62.903134440331506,137.22072841616918,4.996494178803213,-0.3430175693764808,0.011300000000000088,0.0108,0.011800000000000033,846.2222222222222,0.0003404470050428144,0.0003404470050428144,0.0003404470050428144,18\_17:0.009878420587822734; 18\_15:0.014083120311514485; 18\_21:0.0068557125264026575,18\_17:373.3333333333333; 18\_15:412.4444444444444; 18\_21:213.33333333333331
19,146.1898023695175,91.9400226101758,4.784096130303804,-0.3484539565195337,0.009000000000000008,0.0114,0.00660000000000005,846.2222222222222,0.00043798380796432774,0.00043798380796432774,0.00043798380796432774,19\_20:0.015161161729623479; 19\_22:0.009771788738526715; 19\_16:0.006167191319418619,19\_20:302.22222222222223; 19\_22:241.77777777777777; 19\_16:284.44444444444446
20,192.77483373910428,356.2042461762709,5.0423461691617035,-0.342647673737715,0.01749999999999996,0.012599999999999945,0.0224,846.2222222222222,0.00038633921843150636,0.00038633921843150636,0.00038633921843150636,20\_19:0.015161161729623479,20\_19:270.22222222222223;;;;
21,121.18898119765907,162.6862080062389,4.939733852868818,-0.345597545846086,0.016800000000000037,0.016800000000000037,0.0168,846.2222222222222,0.00029414878108628164,0.00029414878108628164,0.00029414878108628164,21\_23:0.004942653566196414; 21\_18:0.0068557125264026575,21\_23:209.77777777777777; 21\_18:277.3333333333333;;
22,179.1917378885148,33.18990160679154,4.725124424572017,-0.3463561119509467,0.015700000000000047,0.01959999999999995,0.0118,846.2222222222222,0.0002420395053297284,0.0002420395053297284,0.0002420395053297284,22\_25:0.007121500396491037; 22\_19:0.009771788738526715,22\_25:344.88888888888886; 22\_19:305.77777777777777;;
23,137.2784109984139,279.49945519940854,4.8048081197662045,-0.34707746460150374,0.008299999999999974,0.0102,0.006399999999999961,846.2222222222222,0.00013277713278296474,0.00013277713278296474,0.00013277713278296474,23\_24:0.0066871824241482025; 23\_21:0.004942653566196414,23\_24:416; 23\_21:273.77777777777777;;
24,126.86824627252496,41.193182296029825,5.0744201617159534,-0.34160804923438676,0.006699999999999928,0.008000000000000007,0.0054,846.2222222222222,0.00036644309698578685,0.00036644309698578685,0.00036644309698578685,24\_23:0.0066871824241482025; 24\_25:0.010029416333921654,24\_23:384; 24\_25:188.44444444444443;;
25,208.7927222869542,168.62567930372066,4.950158025209866,-0.34570711854024605,0.010800000000000032,0.0158,0.005800000000000027,846.2222222222222,0.0001902822073423432,0.0001902822073423432,0.0001902822073423432,25\_22:0.007121500396491037; 25\_24:0.010029416333921654; 25\_26:0.006099378052547205,25\_22:312.88888888888886; 25\_24:252.44444444444443; 25\_26:259.55555555555554
26,108.39938395414997,27.073354718637297,5.151222252121397,-0.3391498922858918,0.00770000000000004,0.0082,0.007199999999999984,846.2222222222222,0.00015230923911980335,0.00015230923911980335,0.00015230923911980335,26\_25:0.006099378052547205,26\_25:195.55555555555554;;;;
\end{filecontents*}
\csvreader[head to column names]{ibmq_ehningen_calibrations_2022-12-02T13_40_10Z.csv}{}
	{\\ \hline \csvcoli & \csvcolii & \csvcoliii & \csvcoliv & \csvcolv & \csvcolvi & \csvcolvii
	& \csvcolviii & \csvcolix & \csvcolx & \csvcolxi & \csvcolxii & \csvcolxiii & \csvcolxiv}
\end{tabular}
}
\caption{Calibration data \texttt{ibmq\_ehningen}.}
\label{table:ehningen}
\end{sidewaystable*}

\begin{sidewaystable*}
\scalebox{0.35}{
\begin{tabular}{|l|l|l|l|l|l|l|l|l|l|l|l|l|l|}%
    \bfseries Qubit & \bfseries T1(us) & \bfseries T2(us) & \bfseries Frequency (GHz) & \bfseries Anharmonicity (GHz) & \bfseries Readout assignment error & \bfseries Prob meas0 prep1 & \bfseries Prob meas1 prep0 & \bfseries Readout length (ns) & \bfseries ID error & \bfseries sx error & \bfseries Pauli-$X$ error & \bfseries CNOT error & \bfseries Gate time (ns)
	\begin{filecontents*}{ibmq_montreal_calibrations_2022-12-02T13_51_28Z.csv}
Qubit,T1 (us),T2 (us),Frequency (GHz),Anharmonicity (GHz),Readout assignment error ,Prob meas0 prep1 ,Prob meas1 prep0 ,Readout length (ns),ID error ,asx (sx) error ,Pauli-X error ,CNOT error ,Gate time (ns);;;;
0,95.62409966137463,139.3509202437932,4.911085728843315,-0.3416299530013241,0.020299999999999985,0.03180000000000005,0.0088,5201.777777777777,0.00021913756648507277,0.00021913756648507277,0.00021913756648507277,0\_1:0.007685720056967521,0\_1:384;;;;
1,142.76177651984966,26.640696530268833,4.834843091816756,-0.32466910427079637,0.017299999999999982,0.0262,0.0084,5201.777777777777,0.00022485221860141715,0.00022485221860141715,0.00022485221860141715,1\_4:0.0079340008655297; 1\_2:0.011943141931053225; 1\_0:0.007685720056967521,1\_4:526.2222222222222; 1\_2:568.8888888888889; 1\_0:419.55555555555554
2,83.35357627492203,118.73367712815882,4.982707339154308,-0.33863810849372844,0.02089999999999992,0.027599999999999958,0.0142,5201.777777777777,0.0003319887992822747,0.0003319887992822747,0.0003319887992822747,2\_3:0.012775122998053168; 2\_1:0.011943141931053225,2\_3:405.3333333333333; 2\_1:533.3333333333333;;
3,93.69674530483532,35.518083226390104,5.105345729994745,-0.3351621911862432,0.02210000000000001,0.03539999999999999,0.0088,5201.777777777777,0.00036137130143264114,0.00036137130143264114,0.00036137130143264114,3\_5:0.00822076311848255; 3\_2:0.012775122998053168,3\_5:398.2222222222222; 3\_2:369.77777777777777;;
4,153.74025904006072,232.0561745114081,5.004009376323411,-0.33873220239666063,0.011399999999999966,0.014399999999999968,0.0084,5201.777777777777,0.00017573715622953072,0.00017573715622953072,0.00017573715622953072,4\_7:0.009852097049896452; 4\_1:0.0079340008655297,4\_7:284.44444444444446; 4\_1:561.7777777777777;;
5,130.99475312437568,132.8479645384429,5.0328798112662145,-0.33720894382745165,0.01869999999999994,0.01980000000000004,0.0176,5201.777777777777,0.0004400550691754118,0.0004400550691754118,0.0004400550691754118,5\_3:0.00822076311848255; 5\_8:0.010007519327498493,5\_3:362.66666666666663; 5\_8:355.55555555555554;;
6,218.96002485268744,30.303425403593874,4.95134349462269,-0.4198945429033115,0.1059000000000001,0.051,0.16080000000000005,5201.777777777777,0.00048070926516418527,0.00048070926516418527,0.00048070926516418527,6\_7:0.013391891277481915,6\_7:526.2222222222222;;;;
7,124.73036558920737,130.05816219810092,4.902297686735172,-0.3233875324821201,0.08420000000000005,0.046,0.12239999999999995,5201.777777777777,0.000289797938134072,0.000289797938134072,0.000289797938134072,7\_10:0.017932077946633568; 7\_6:0.013391891277481915; 7\_4:0.009852097049896452,7\_10:1059.5555555555554; 7\_6:490.66666666666663; 7\_4:320
8,108.2854413980355,125.2651276041562,4.908075761447362,-0.3242773911866739,0.01739999999999997,0.023599999999999954,0.0112,5201.777777777777,0.0001846673668748441,0.0001846673668748441,0.0001846673668748441,8\_9:0.006734509404882627; 8\_11:0.00835632896362179; 8\_5:0.010007519327498493,8\_9:405.3333333333333; 8\_11:483.55555555555554; 8\_5:391.1111111111111
9,154.2890986539428,162.7392707117348,5.0447386108364665,-0.33853943712278195,0.011600000000000055,0.0194,0.0038000000000000256,5201.777777777777,0.00042817392085195785,0.00042817392085195785,0.00042817392085195785,9\_8:0.006734509404882627,9\_8:369.77777777777777;;;;
10,110.00565766273078,166.43765680954846,5.081902606081007,-0.3379636840612414,0.010199999999999987,0.011600000000000055,0.0088,5201.777777777777,0.0003708966208295047,0.0003708966208295047,0.0003708966208295047,10\_7:0.017932077946633568; 10\_12:0.011379294517449401,10\_7:1024; 10\_12:412.4444444444444;;
11,104.36466512685381,55.7591311324293,5.0348241050985765,-0.3377232857931562,0.012799999999999923,0.019399999999999973,0.0062,5201.777777777777,0.00039356128747016113,0.00039356128747016113,0.00039356128747016113,11\_14:0.0060915490158756636; 11\_8:0.00835632896362179,11\_14:341.3333333333333; 11\_8:448;;
12,168.95389936976417,158.22763043069898,4.971885530892281,-0.3227716338059607,0.01539999999999997,0.023599999999999954,0.0072,5201.777777777777,0.00029662115143645545,0.00029662115143645545,0.00029662115143645545,12\_15:0.011652730322951027; 12\_13:0.008910570465851392; 12\_10:0.011379294517449401,12\_15:405.3333333333333; 12\_13:391.1111111111111; 12\_10:376.88888888888886
13,185.7212041220119,120.9249726699928,4.8677599397255,-0.3406276441736398,0.014100000000000001,0.020199999999999996,0.008,5201.777777777777,0.0003167504377956908,0.0003167504377956908,0.0003167504377956908,13\_12:0.008910570465851392; 13\_14:0.010307223360133233,13\_12:426.66666666666663; 13\_14:490.66666666666663;;
14,102.86372644261384,156.15806787132806,4.961377834548816,-0.32314254463107384,0.06129999999999991,0.11419999999999997,0.0084,5201.777777777777,0.00047655217400880925,0.00047655217400880925,0.00047655217400880925,14\_16:0.008808088995997276; 14\_11:0.0060915490158756636; 14\_13:0.010307223360133233,14\_16:355.55555555555554; 14\_11:376.88888888888886; 14\_13:526.2222222222222
15,134.13429656078398,105.30324210293969,5.033922349121523,-0.3374199469605722,0.05120000000000002,0.0928,0.0096,5201.777777777777,0.0007762286408396536,0.0007762286408396536,0.0007762286408396536,15\_18:0.01424504237198776; 15\_12:0.011652730322951027,15\_18:597.3333333333333; 15\_12:369.77777777777777;;
16,125.09794331057523,90.7427647738559,5.08571624759576,-0.3370728696076825,0.009199999999999986,0.01419999999999999,0.0042,5201.777777777777,0.00022580879597816354,0.00022580879597816354,0.00022580879597816354,16\_19:0.012207688072333495; 16\_14:0.008808088995997276,16\_19:270.22222222222223; 16\_14:320;;
17,107.57414064726706,112.19082777747501,5.072203793541216,-0.3376263126986256,0.038899999999999935,0.045399999999999996,0.0324,5201.777777777777,0.0004030689297348349,0.0004030689297348349,0.0004030689297348349,17\_18:0.010442988159402977,17\_18:348.4444444444444;;;;
18,58.05184326230592,19.01296244353661,4.980935429851672,-0.3038509594814633,0.03520000000000001,0.04920000000000002,0.0212,5201.777777777777,0.00041849336532723416,0.00041849336532723416,0.00041849336532723416,18\_21:0.011902961885060848; 18\_15:0.01424504237198776; 18\_17:0.010442988159402977,18\_21:440.88888888888886; 18\_15:632.8888888888888; 18\_17:384
19,106.65699448678816,155.73366062530962,4.981775368607511,-0.32218874979120343,0.012899999999999912,0.018399999999999972,0.0074,5201.777777777777,0.0003181069743700482,0.0003181069743700482,0.0003181069743700482,19\_22:0.007166386822440668; 19\_16:0.012207688072333495; 19\_20:0.006643091214783126,19\_22:327.1111111111111; 19\_16:305.77777777777777; 19\_20:355.55555555555554
20,129.7723268212009,148.17424191213706,5.082266923091465,-0.3372175995104487,0.010999999999999899,0.016599999999999948,0.0054,5201.777777777777,0.00020752965693358994,0.00020752965693358994,0.00020752965693358994,20\_19:0.006643091214783126,20\_19:320;;;;
21,96.89151383151487,57.66352184951713,5.0935872122178525,-0.3358012541809501,0.046499999999999986,0.024399999999999977,0.0686,5201.777777777777,0.00042578256353051407,0.00042578256353051407,0.00042578256353051407,21\_18:0.011902961885060848; 21\_23:0.010028945656124888,21\_18:405.3333333333333; 21\_23:426.66666666666663;;
22,97.05813793076156,191.12699519941407,5.056663366175174,-0.33710426431518753,0.01629999999999998,0.025000000000000022,0.0076,5201.777777777777,0.00026125245199208483,0.00026125245199208483,0.00026125245199208483,22\_19:0.007166386822440668; 22\_25:0.008099066232685859,22\_19:291.55555555555554; 22\_25:611.5555555555555;;
23,162.49733002220748,35.63487116109012,4.972903367566311,-0.337396806334524,0.01319999999999999,0.02059999999999995,0.0058,5201.777777777777,0.0003254527052676989,0.0003254527052676989,0.0003254527052676989,23\_24:0.007677210047334987; 23\_21:0.010028945656124888,23\_24:405.3333333333333; 23\_21:391.1111111111111;;
24,78.75943964920076,36.49145558812868,5.05141613198169,-0.33866209517065726,0.020999999999999908,0.023399999999999976,0.0186,5201.777777777777,0.0003248239445235739,0.0003248239445235739,0.0003248239445235739,24\_23:0.007677210047334987; 24\_25:0.010109435718178184,24\_23:440.88888888888886; 24\_25:376.88888888888886;;
25,90.64944244508125,113.0138698526409,4.934830844142309,-0.3225452402104976,0.01639999999999997,0.0232,0.0096,5201.777777777777,0.00021398760195830708,0.00021398760195830708,0.00021398760195830708,25\_24:0.010109435718178184; 25\_26:0.014343313245431616; 25\_22:0.008099066232685859,25\_24:412.4444444444444; 25\_26:483.55555555555554; 25\_22:576
26,128.9370805380526,164.56819759243487,4.99967547641022,-0.3377513287045508,0.014799999999999924,0.021599999999999953,0.008,5201.777777777777,0.0005686469060359692,0.0005686469060359692,0.0005686469060359692,26\_25:0.014343313245431616,26\_25:519.1111111111111;;;;
\end{filecontents*}
\csvreader[head to column names]{ibmq_montreal_calibrations_2022-12-02T13_51_28Z.csv}{}
	{\\ \hline \csvcoli & \csvcolii & \csvcoliii & \csvcoliv & \csvcolv & \csvcolvi & \csvcolvii
	& \csvcolviii & \csvcolix & \csvcolx & \csvcolxi & \csvcolxii & \csvcolxiii & \csvcolxiv}
\end{tabular}
}
\caption{Calibration data \texttt{ibmq\_montreal}.}
\label{table:montreal}
\end{sidewaystable*}

\end{document}